\documentclass[journal]{IEEEtran}
\usepackage{color,soul}
\usepackage{graphicx}
\usepackage{float}
\graphicspath{{figs/}}
\usepackage{amsmath}
\usepackage{amssymb, enumitem}
\usepackage{multirow}

\DeclareMathOperator{\E}{\mathbb{E}}

\begin{document}

%\title{Self Attention guided Synthesis and Confidence based Segmentation of Brain Anatomy from 2D Ultrasound}

\title{Learning to Segment Brain Anatomy from 2D Ultrasound with Less Data}

\author{Jeya~Maria~Jose~V.,~\IEEEmembership{Student Member,~IEEE,}
        Rajeev Yasarla,~\IEEEmembership{Student Member,~IEEE,}, Puyang Wang,~\IEEEmembership{Student Member,~IEEE,}, Ilker Hacihaliloglu,~\IEEEmembership{Member,~IEEE,} and Vishal~M.~Patel,~\IEEEmembership{Senior Member~,~IEEE}% <-this % stops a space
\thanks{Jeya Maria Jose V., is with the Whiting School of Engineering, Johns Hopkins University, 3400 North Charles Street, Baltimore, MD 21218-2608, e-mail: jvalana1@jhu.edu
	\footnotesize}
\thanks{Rajeev Yasarla, is with the Whiting School of Engineering, Johns Hopkins University, 3400 North Charles Street, Baltimore, MD 21218-2608, e-mail: ryasarl1@jhu.edu}% <-this % stops a space
\thanks{Puyang Wang, is with the Whiting School of Engineering, Johns Hopkins University, 3400 North Charles Street, Baltimore, MD 21218-2608, e-mail: pwang47@jhu.edu }
\thanks{Ilker Hacihaliloglu, is with the Department of Biomedical Engineering, Rutgers, The State University of New Jersey, e-mail: ilker.hac@soe.rutgers.edu }
\thanks{Vishal M. Patel, is with the Whiting School of Engineering, Johns Hopkins University, e-mail: vpatel36@jhu.edu}
\thanks{Manuscript received...}}

\maketitle

\begin{abstract}
Automatic segmentation of anatomical landmarks from ultrasound (US) plays an important role in the management of preterm neonates with a very low birth weight due to the increased risk of developing intraventricular hemorrhage (IVH) or other complications. One major problem in developing an automatic segmentation method for this task is the limited availability of annotated data. To tackle this issue, we propose a novel image synthesis method using multi-scale self attention generator to synthesize US images from various segmentation masks.  We show that our method can synthesize high-quality US images for every manipulated segmentation label with qualitative and quantitative improvements over the recent state-of-the-art  synthesis methods.
Furthermore, for the segmentation task, we propose a novel method, called Confidence-guided Brain Anatomy Segmentation (CBAS) network, where segmentation and corresponding confidence maps are estimated at different scales. In addition, we introduce a technique which guides  CBAS to learn the weights based on the confidence measure about the estimate. Extensive experiments demonstrate that the proposed method for both synthesis and segmentation tasks achieve significant improvements over the recent state-of-the-art methods.  In particular, we show that the new synthesis framework can be used to generate realistic US images which can be used to improve the performance of a segmentation algorithm.
\end{abstract}

% Note that keywords are not normally used for peerreview papers.

\IEEEpeerreviewmaketitle

\section{Introduction}
According to the World Health Organization, 15 million babies are born preterm each year \cite{blencowe2013born}. Although, advancements made in neonatal care have increased the survival rates, majority of these infants are at risk for long-term complications such as cerebral palsy, cognitive-behavioral and learning impairments. In premature infants, one of the most common brain injury is intraventricular hemorrhage (IVH) \cite{robinson2012neonatal}. These hemorrhages result in ventricle dilation, which can lead to serious brain damage if not properly treated.
Ventricle dilation is also associated with white matter atrophy (hydrocephalus ex-vacuo). Therefore, monitoring of ventricle volume change in neonates is clinically important in order to determine the correct intervention. On the other hand  absence of septum pellucidum is used as a valuable landmark for the diagnosis of abnormalities, such as septo-optic dysplasia, in the central nervous system (CNS) \cite{sherer2004prenatal,sarwar1989septum}. The main imaging modality currently employed for monitoring brain abnormalities in preterm neonates is two-dimensional (2D) ultrasound (US) due to its real-time safe imaging capabilities. However, high levels of noise and various imaging artifacts, and irregular shape deformation of ventricles, results in the inability to localize the site and extent of brain injury, or to predict neurologic outcomes in identifying IVH or other abnormalities from US data. Being a user dependent imaging modality causes additional difficulties during data collection since a single-degree deviation angle by the operator can reduce the signal strength by 50\%. Current clinical practice involves manual measurement of ventricle or investigation of septum pellucidum presence from the collected scans by clinicians. Due to previously mentioned difficulties, related to US imaging, this is an error prone and time consuming process.

\begin{figure}[t!]
	\centering
	\includegraphics[width=0.15\textwidth,height = 0.15\textwidth]{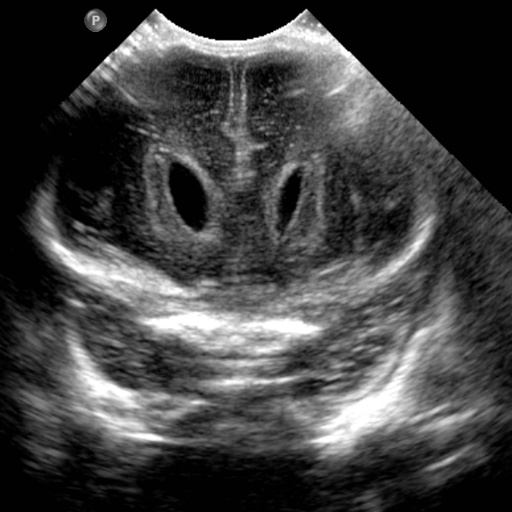}
	\includegraphics[width=0.15\textwidth,height = 0.15\textwidth]{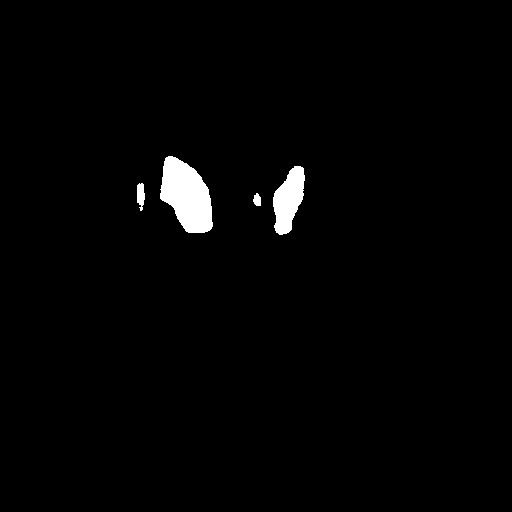}
	\includegraphics[width=0.15\textwidth,height = 0.15\textwidth]{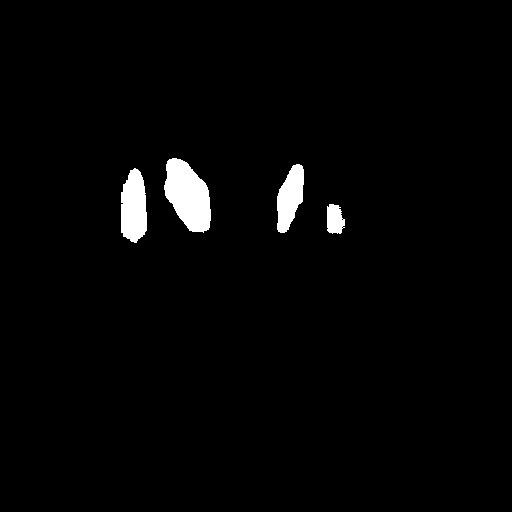}\\
	(a)\hskip70pt (b)\hskip70pt (c)\\
	\includegraphics[width=0.15\textwidth,height = 0.15\textwidth]{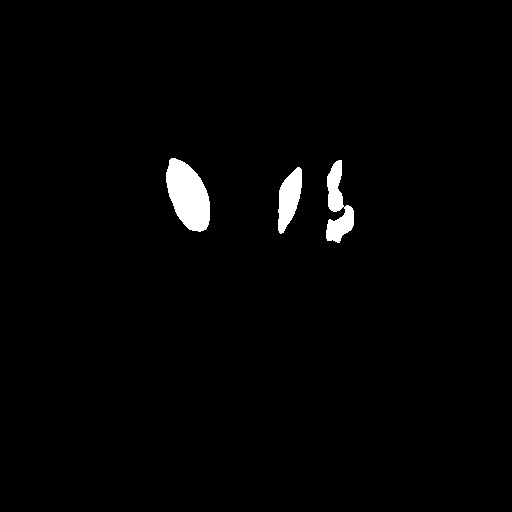}
	\includegraphics[width=0.15\textwidth,height = 0.15\textwidth]{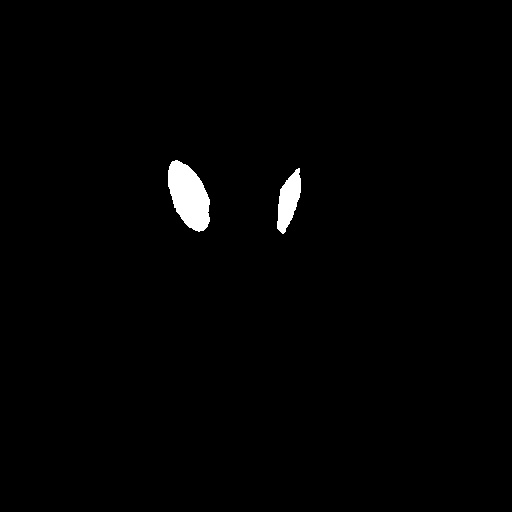}
	\includegraphics[width=0.15\textwidth,height = 0.15\textwidth]{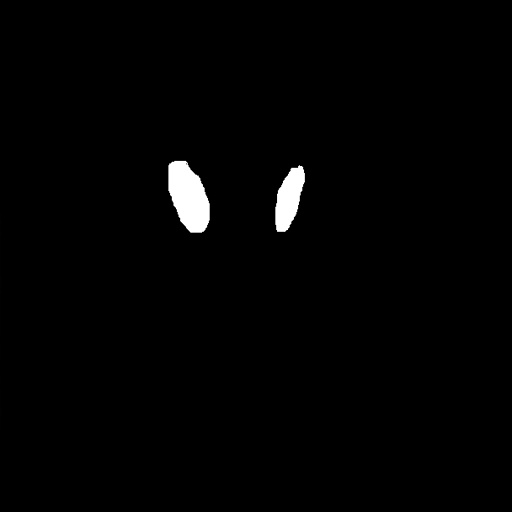}\\
	(d)\hskip70pt (e)\hskip70pt (f)\\
	\caption{(a) Original brain US image. Brain ventricular segmentation obtained using (b) pix2pix \cite{isola2017image}, (c) U-net \cite{unet}, (d) Wang et al.\cite{wang2018automatic}, (e) CBAS (ours). (f) ground-truth brain ventricular regions.}
	\label{Fig:exp_in}
\end{figure}

\begin{figure*}[t!]
	\begin{center}
		\centering
		\includegraphics[width=0.18\textwidth,height = 0.15\textwidth]{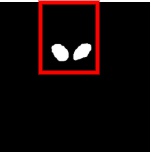}
		\includegraphics[width=0.18\textwidth,height = 0.15\textwidth]{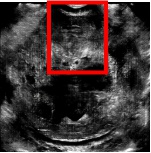}
		\includegraphics[width=0.18\textwidth,height = 0.15\textwidth]{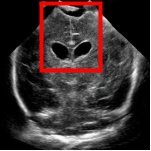}
		\includegraphics[width=0.18\textwidth,height = 0.15\textwidth]{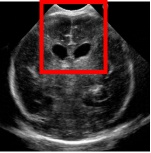}
		\includegraphics[width=0.18\textwidth,height = 0.15\textwidth]{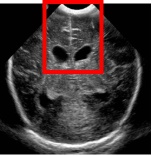}\\ \vskip2pt
		\includegraphics[width=0.18\textwidth,height = 0.15\textwidth]{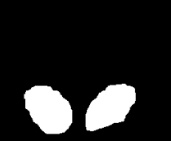}
		\includegraphics[width=0.18\textwidth,height = 0.15\textwidth]{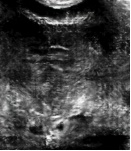}
		\includegraphics[width=0.18\textwidth,height = 0.15\textwidth]{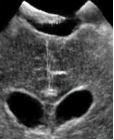}
		\includegraphics[width=0.18\textwidth,height = 0.15\textwidth]{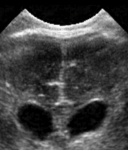}
		\includegraphics[width=0.18\textwidth,height = 0.15\textwidth]{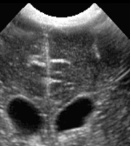}\\
		(a)\hskip85pt(b)\hskip85pt(c)\hskip85pt(d)\hskip85pt(e)
		\caption{(a) Input segmentation mask. Synthesized image using (b) pix2pix \cite{isola2017image} (c) pix2pixHD \cite{wang2018high} (d) MSSA (ours) (e) Original image corresponding to the segmentation mask in (a). The second row consists of the zoomed portions of the image inside the red box in the first row.}
		\label{Fig:synex}
	\end{center}
\end{figure*}

In order to automate the ventricle segmentation and measurement process, various groups have proposed automatic segmentation methods. In \cite{qiu2017automatic}, a fully automated atlas-based segmentation pipeline was developed for segmenting 3D volumetric US data. Validation results performed on 30 3D US scans, obtained from 14 patients, achieved a mean Dice similarity coefficient (DSC) and maximum absolute distance of $76.5\%$ and 1 mm, respectively. The reported computation time for segmenting a single 3D volume was 54 mins \cite{qiu2017automatic}. Atlas-based volumetric US segmentation was also proposed in \cite{boucher2018dilatation}. Validation performed on 16 subject scans achieved a mean DSC of 0.70. Computation time was not reported. A semi-automatic method, for segmenting volumetric US scans, was proposed in \cite{sciolla2016segmentation}. Mean absolute distance between the manual and semi-automatically segmented contours was 2.17 mm. Subject size and processing time was not reported \cite{sciolla2016segmentation}. In order to improve the accuracy and computation time, methods based on deep learning have been investigated \cite{martin2018automatic,wang2018automatic}. In \cite{martin2018automatic}, a U-net based \cite{unet} network architecture was proposed. Reported mean DSC value and computation time were 0.81 and 5 seconds per volume (0.01 seconds per slice) respectively for 15 volumes obtained from 14 patients. In \cite{wang2018automatic}, a multi-scale-based network architecture was proposed for segmentation of 2D US scans. Validation studies performed on 687 scans, obtained from 10 subjects, achieved a mean DSC value of 0.90 with a computation time of 0.02 seconds.

Although, deep learning methods have resulted in increased accuracy and computation time, most of the previous work has been validated on scans with enlarged ventricles. If the foreground anatomical structure,  anatomy to be segmented, is significantly smaller compared to the background anatomical structure, traditional convolutional neural network (CNN) architectures fail since there is not enough positional information to localize small brain anatomy. The same is also valid for segmenting densely packed small brain anatomy (small ventricles and septum pellecudi appearing in the middle of the US scan). Finding small anatomical structure using a CNN architecture is difficult since  resolution of small features is gradually lost and resulting coarse features can miss the details of small structures \cite{hamaguchi2018effective}. For example, methods like pix2pix \cite{isola2017image}, U-net \cite{unet}, and Wang et al. \cite{wang2018automatic} fail to segment the brain ventricular region from the US images as shown in Fig.~\ref{Fig:exp_in}. These methods end up segmenting the non-ventricular region as the brain ventricular region.  This is mainly due to the lack of special attention given to small ventricles while learning the network weights. Finally, due to the high complexity and variability in the ventricles shape, the traditional CNN architectures result in over or under segmentation (Fig.~\ref{Fig:exp_in}).  

To address this problem, we propose a method called, Confidence-guided Brain Anatomy Segmentation (CBAS) network, where we make use of the aleatoric uncertainty and define confidence scores at each pixel which are data dependent. Uncertainty can be modeled in two ways -- epistemic and aleatoric uncertainties as explained in \cite{KendallGal2017UncertaintiesB,kendall2017multi}. In order to achieve better performance in  tasks like medical image segmentation, \cite{mehta2019propagating,nair2020exploring,jungo2019assessing} modeled epistemic uncertainty for learning the CNN network weights. To handle different brain anatomy structures in 2D US scans, we define data dependent aleatoric uncertainty as the confidence scores that are computed by the confidence blocks in CBAS.  These blocks essentially  indicate how confident the CBAS network is about the segmentation output. This confidence score will be low for the regions where the error is high and vice-versa. Thus CBAS learns to differentiate the erroneous regions and gives special attention to those regions in subsequent layers while computing the segmentation output.  We present a novel method for fully automatic ventricles and septum pellecudi segmentation with varying size from 2D US scans. Note that this is the first approach that uses uncertainty in 2D US segmentation. We validate our method on 1629 US scans obtained from 20 different subjects.

One of the major problems in medical image analysis is the limited number of annotated data. Obtaining clinical annotations is also a difficult, expensive and time consuming process as expert radiologists are needed. For very specific tasks like the one addressed in this paper, the availability of datasets is also very scarce. As most of the current state-of-the-art segmentation methods require a considerable amount of data to train the network, using them for tasks with less data does not guarantee a good performance.  As a result, novel image synthesis methods are proposed in the literature to syntehsize meaningful high quality data that could be added to the training dataset.

Over the past few years, image synthesis and image-to-image translation tasks have been dominated by Generative Adversarial Networks (GANs) \cite{goodfellow2014generative} and its variations. In this approach, a generator is trained to synthesize an image from random noise while a discriminator, which is trained on both real and synthesized images tries to classify whether the image is real or was synthesized by the generator (i.e. fake). Both networks are trained in a min-max way such that they act as adversaries of each other. While using GANs in medical imaging to synthesize new images solves the issue of limited availability of data, the problem of annotations still exists in this setup. Isola et al. \cite{isola2017image} proposed using a conditional generative adversarial network (cGAN) \cite{mirza2014conditional} to solve the image-to-image translation tasks where the network is trained to learn the mapping between an image across two different domains. In the medical imaging community, several works (\cite{nie2018medical}, \cite{nie2017medical}, \cite{wolterink2017deep}, \cite{bi2017synthesis}, \cite{han2018gan}, \cite{yang2018mri}, \cite{armanious2019medgan}) have adapted this idea to synthesize images from one modality to another modality such as MRI to CT, T1 MRI to T2 MRI etc.  Since this method can be used for any translation task, it can be used for image synthesis from segmentation labels where the network is trained to translate the segmentation mask of an image into a realistic US image. Zhao et al. \cite{zhao2018synthesizing} showed that multiple realistic-looking retinal images can be synthesized from just the annotation masks using this method. Bailo et al. \cite{bailo2019red} used cGAN to generate blood smear image data from segmentation masks corresponding to microscopic images. Diverse set of new images were also achieved by manipulating the segmentation labels. Jaiswal et al. \cite{jaiswal2018capsulegan} used a capsule cGAN to synthesize microscopic data of cortical axons.

\begin{figure*}[t!]
	\begin{center}
		\centering
		\includegraphics[width=0.8\textwidth]{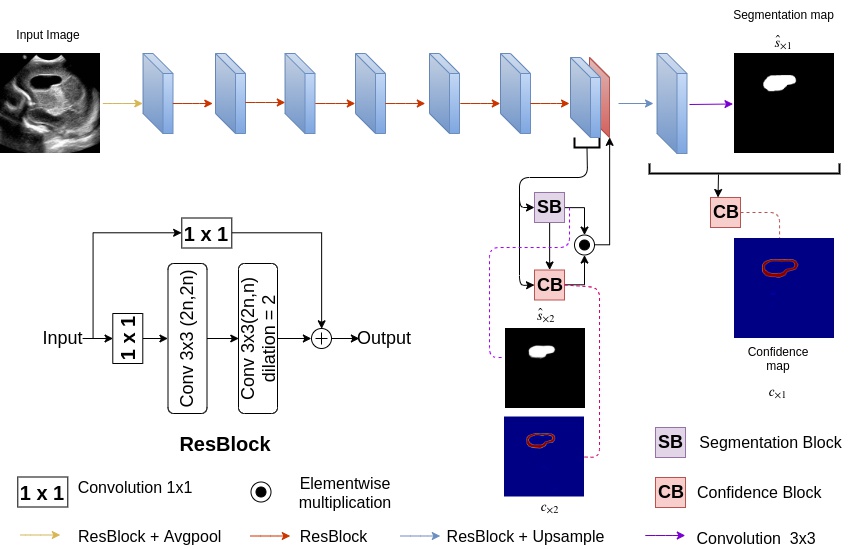}
		\caption{An overview of the proposed CBAS network.  The aim of the CBAS network is to estimate the brain anatomy segmentation for the given brain US image. CBAS learns the segmentation maps and computes the confidence maps to guide the network. To achieve this, we introduce SB and CB networks and feed their outputs to the subsequent layers. Note that in the confidence maps blue means 1 and red means 0}
		\label{Fig:CBVS}
	\end{center}
\end{figure*}

Though many methods exist for medical image synthesis, most of them only deal with generating low resolution images. Synthesis of US images is harder compared to other imaging modalities as it contains speckle and many artifacts when compared to other imagining modalities. In Fujioka et al. \cite{fujioka2019breast} breast US images were synthesized using a GAN-based approach \cite{radford2015unsupervised}. The authors however fail to show any quantitative analysis or usefulness of the synthesized images. In \cite{hu2017freehand}, fetal US images were synthesized from tracked B-mode US data. Validation experiments were performed on US data obtained from a fetus ultrasound examination phantom. Simulation of realistic in vivo US data is a more challenging problem as soft tissue properties vary significantly depending on the imaged subject and orientation of the transducer \cite{hu2017freehand}. In \cite{tom2018simulating}, GANs were used to simulate intravascular US (IVUS) data using convolution networks. Most of the synthesis methods use convolutional networks fail to capture the long range dependencies in the image due to the low receptive field of convolution. This can be clearly seen by comparing the performance of different synthesis methods as shown in Fig. \ref{Fig:synex}.  It can be observed that pix2pix \cite{isola2017image} synthesizes very poor quality image and pix2pixHD \cite{wang2018high} fails to capture the fine details of the ultrasound image towards the edges. To tackle this issue, we propose a novel attention-based method that can synthesize realistic brain US images from a ventricle and septum pellecudi segmentation masks.  We use a multi-scale generator architecture with multi-scale self-attention modules that guides the network to capture the long range dependencies while also synthesizing high resolution images.  A sample synthesized image using our method is shown in Fig~\ref{Fig:synex}(d).  As compared to the other synthesis methods, the proposed method produces sharper images from the input segmentation masks.   Using the proposed synthesis model, numerous realistic US images can be synthesized by manipulating the segmentation masks that is fed into the network. As the images are directly synthesized from the manipulated segmentation masks, there is no need for annotation of the synthesized data. By performing extensive experiments, we show that these synthesized images, when added to the training data, increase the performance of the segmentation network.

This paper makes the following contributions:
\begin{itemize}
	\item A novel synthesis network is proposed using a multi-scale generator guided by self-attention modules to synthesize realistic US images from the segmentation masks.
	\item A novel US image segmentation method, called CBAS, is proposed which generates the segmentation maps at different scales along with the confidence maps, to guide subsequent layers the network by blocking the propagation of errors in the segmentation map at lower scale, while computing final output segmentation.
	\item A novel loss function is introduced to train CBAS which makes use of the computed confidence maps and the corresponding segmentation maps.
	\item Extensive experiments are conducted to show the significance of the proposed synthesis and segmentation networks.   Furthermore, an ablation study is conducted to demonstrate the effectiveness of different parts of our networks. We also show that the synthesized images are useful as they can be used to improve the segmentation performance.
\end{itemize}

Rest of the paper is organized as follows. Details of the proposed uncertainty-guided segmentation method are given in Section~\ref{sec:propose}.  Section~\ref{sec:synthesis} gives details regarding the proposed self-attention based synthesis method.  Experimental results as well as ablation study details are given in Section~\ref{exp_res}.  Finally, Section~\ref{conc} concludes the paper with a brief summary and discussion.

\section{Confidence-guided Brain Anatomy Segmentation (CBAS)}
\label{sec:propose}

Let the set of brain US scans be denoted as $\mathcal{B}$ and the corresponding set of brain ventricle segmentation maps as $\mathcal{S}$. Our aim is to estimate the brain ventricle segmentation map $\hat{s}$ for a given brain US scan $x\in\mathcal{B}$. To address this probelm unlike many deep learning-based methods that directly estimate the brain ventricle segmentation map, we take a different approach in which we first estimate the segmentation map $s$ and the corresponding confidence map $c$. We define the confidence map $c$, that represents the confidence score at each pixel which resembles the measure of how much the network is certain about the computed value in the segmentation map. Our proposed method, CBAS, judiciously combines the segmentation and confidence information at lower scales to block the propagation of errors in the segmentation $\hat{s}_{\times 2}$ while computing the final segmentation map $\hat{s}$.   Fig.~\ref{Fig:CBVS} gives an overview of the proposed CBAS network.  As can be seen from this figure, we estimate the segmentation map $\hat{s}_{\times 2}$ and the confidence map $c_{\times 2}$ at scale $\times 2$ (0.5 scale of $x$) and they are fed back to the subsequent layers in a way that blocks the errors in $\hat{s}_{\times 2}$ using $c_{\times 2}$.

In CBAS, we estimate the segmentation maps at two different scales, $i \in \{\times 1,\times 2\}$, i.e $\hat{s}_{\times 1}$ (same size as $x$) and $\hat{s}_{\times 2}$ (0.5 scale as $x$), and the corresponding confidence maps $c_{\times 1}$ and $c_{\times 2}$. To estimate these segmentation maps, we construct our base network (BN) using UNet \cite{unet} architecture with the ResBlock as our basic building block. To increase the receptive field size, we introduce dilation convolutions in the ResBlock, as shown in Fig.~\ref{Fig:CBVS}, where Conv~$l\times~l$~($m,n$) contains instance normalization \cite{Instance2016}, Rectified Linear Unit (ReLU), Conv ($l \times l$) - convolutional layer with kernel of size $l \times l$, where $m$ and $n$ are the number of input and output channels, respectively. Note that all convolutional layers in BN are densely connected \cite{huang2017densely}. The BN network consists of the following sequence of layers:\\
ResBlock(1,32)-Avgpool-ResBlock(32,32)-Avgpool-ResBlock(32,32)-\\
ResBlock(32,32)-ResBlock(32,32)-ResBlock(32,32)-Upsample-ResBlock(32,32)-\\Upsample-ResBlock(33,16)-Conv$ 3\times 3$(16,1),\\
where Avgpool is the average pooling layer, and Upsample is the upsampling convolution layer.

\begin{figure*}[t!]
	\begin{center}
		\centering
		\includegraphics[width=\textwidth]{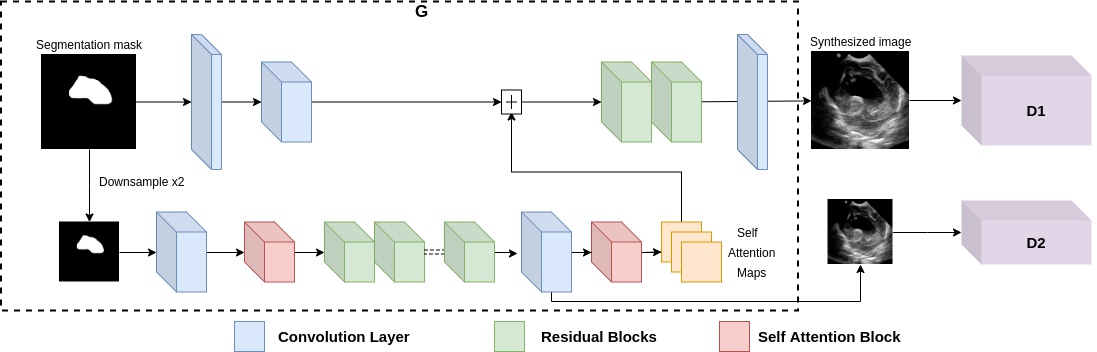}
		\caption{An overview of the proposed MSSA network.  The MSSA network takes in a segmentation mask and synthesizes the corresponding realistic looking synthetic ultrasound image. G denotes the generator. D1 and D2 denote the discriminators across each scale.}
		\label{Fig:MSSA}
	\end{center}
\end{figure*}

\subsection{CBAS Network}
Segmentation networks are prone to misclassify the labels at or near the edges of brain ventricles. Hence a brain ventricle segmentation method requires a special attention in those regions where the network may go wrong. To address this issue, one can estimate brain ventricle segmentation at different scales, and estimate the confidence map which indicates the regions where the method can go wrong. Confidence map highlights the regions where the network is certain about the segmentation values by producing high confidence values (i.e nearly 1) and assigning low confidence scores for those pixels where the network is uncertain. In this way, highlighting the regions in the confidence map and combing them with the segmentation map, we block the propagation of errors in segmentation, and make the network more attentive in the erroneous regions. To estimate these pairs of segmentation and the corresponding confidence map, we introduce Segmentation Block (SB) and Confidence Block (CB) in our base network (BN) and construct our CBAS network as shown in Fig.~\ref{Fig:CBVS}.

\subsection{Segmentation and Confidence Blocks}  
Feature maps at scale $\times 2$ are give as input to the Segmentation Block (SB) to compute the segmentation map $s_{\times 2}$. SB is a sequence of four convolutional layers.  We feed the estimated segmentation maps and the feature maps as inputs to CB for computing the confidence score at every pixel, which indicates how  certain the network is about the segmentation value. CB is a sequence of four convolutional layers.  Details of convolutional layers in SB and CB blocks  are shown in Table~\ref{block_tables} (in Appendix A). 

Given an US image $x$, we estimate the segmentation maps ($\hat{s}_{\times 1}$ and $\hat{s}_{\times 2}$) as well as the corresponding confidence maps ($c_{\times 1}$ and $c_{\times 2}$) as shown in Fig.~\ref{Fig:CBVS}. We propose a confidence-guided loss function to train the CBAS network which uses a pair of segmentation and confidence maps (i.e \{$\hat{s}_{\times 1},\; c_{\times 1}$\} and \{$\hat{s}_{\times 2},\; c_{\times 2}$\}).

\subsection{Loss for CBAS}
We use the confidence to guide the CBAS network in learning the weights of the network. We define the confidence guided loss as,
\begin{equation}
\mathcal{L}_{final} = \sum_{i\in \{\times 1,\:\times 2\}}\sum_{j}\sum_{k}c_{i_{jk}}\mathcal{L}_{CE}(\hat{s}_{i_{jk}},s_{i_{jk}}) -\lambda \log(c_{i_{jk}}),
\label{Eq:eq5}
\end{equation}
where $$\mathcal{L}_{CE}(\hat{s}_{jk},s_{jk}) = -s_{jk}\log(\hat{s}_{jk})-(1-s_{jk})\log(1-\hat{s}_{jk}),$$ and $\lambda$ is a constant.

Inspired by the aleatoric uncertainty \cite{KendallGal2017UncertaintiesB,kendall2017multi}, in our method, we attempt to address the data dependent uncertainty caused in the outputs due to different sizes of the brain ventricles and sensor noise which are inherent in US brain images. We formulate this data dependent uncertainty as the confidence score ($c$), i.e finding a confidence score at every pixel in the output which depends on the input brain US scan. We compute these confidence scores using CB (confidence block) as explained in the earlier section. Computing these confidence scores benefits us in learning the network weights as the erroneous regions have low confidence scores. Note that, to capture the erroneous regions, the confidence score should be estimated pixel-wise. It is beneficial to guide the network by recognizing the regions which are prone to make errors if we estimate the confidence scores pixel-wise. Note that values in the confidence map at every position will be in the range of $[0,1]$.

\begin{figure*}[htp!]
	\begin{center}
		\centering
		\includegraphics[width=0.13\textwidth,height = 0.13\textwidth]{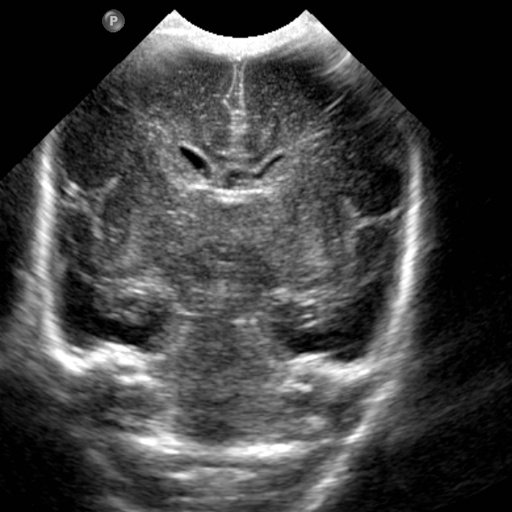}
		\includegraphics[width=0.13\textwidth,height = 0.13\textwidth]{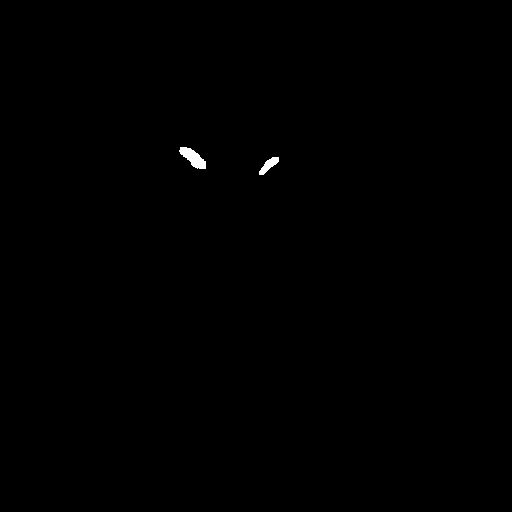}
		\includegraphics[width=0.13\textwidth,height = 0.13\textwidth]{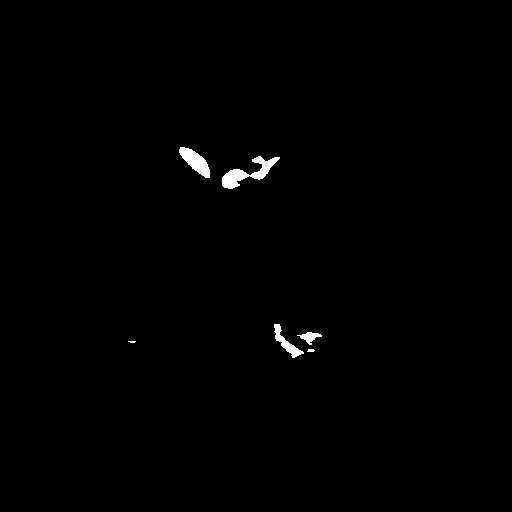}
		\includegraphics[width=0.13\textwidth,height = 0.13\textwidth]{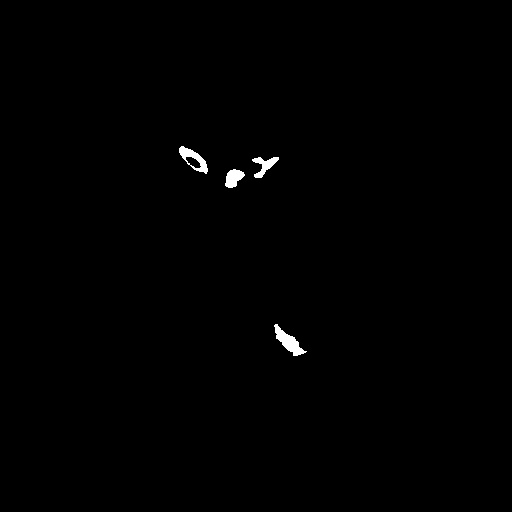}
		\includegraphics[width=0.13\textwidth,height = 0.13\textwidth]{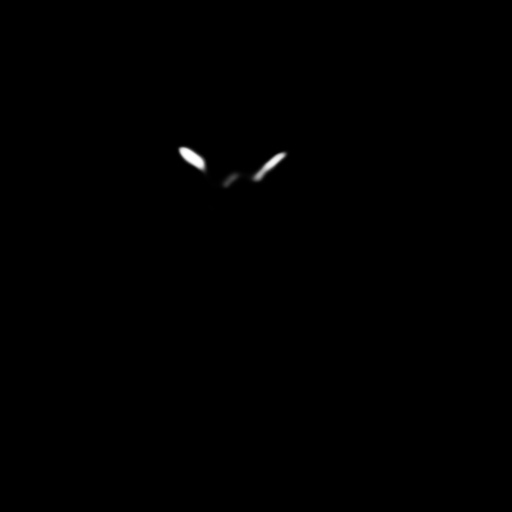}
		\includegraphics[width=0.13\textwidth,height = 0.13\textwidth]{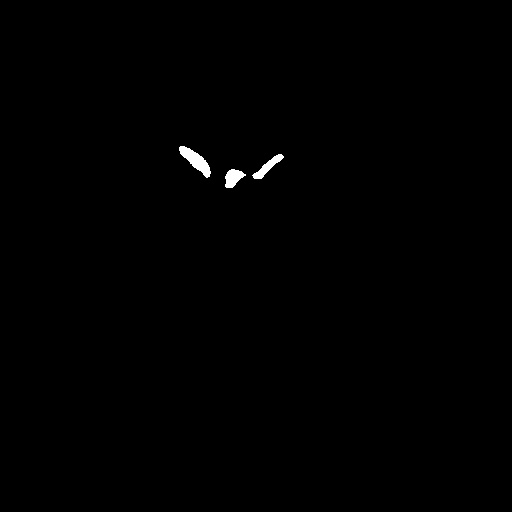}
		\includegraphics[width=0.13\textwidth,height = 0.13\textwidth]{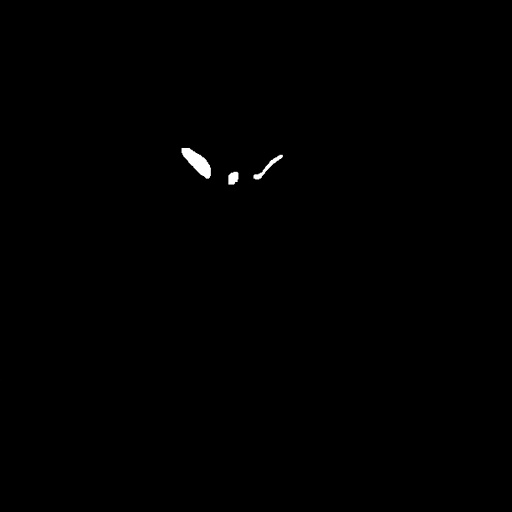}\\
		\vskip4pt
		\includegraphics[width=0.13\textwidth,height = 0.13\textwidth]{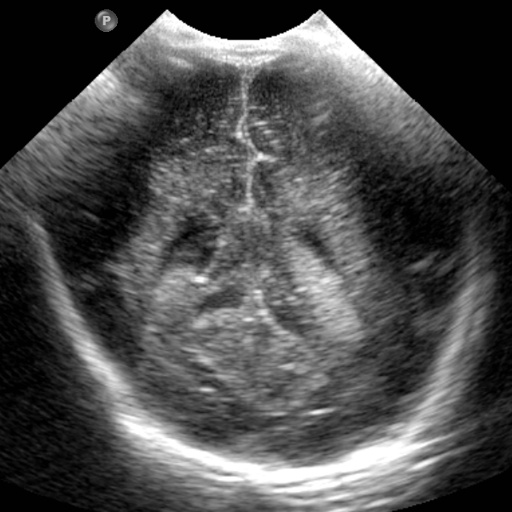}
		\includegraphics[width=0.13\textwidth,height = 0.13\textwidth]{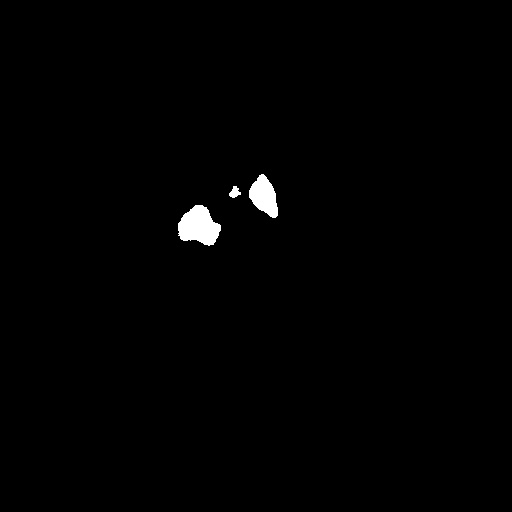}
		\includegraphics[width=0.13\textwidth,height = 0.13\textwidth]{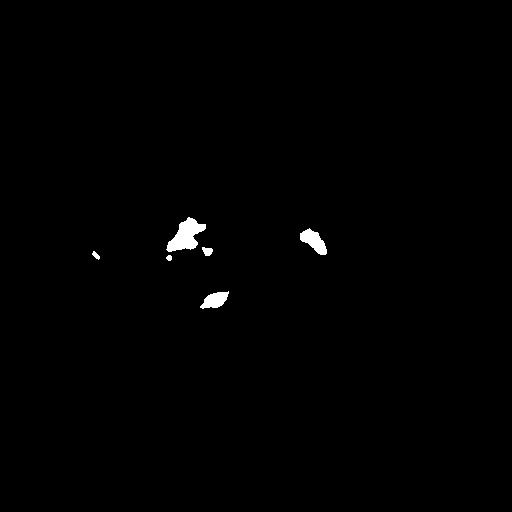}
		\includegraphics[width=0.13\textwidth,height = 0.13\textwidth]{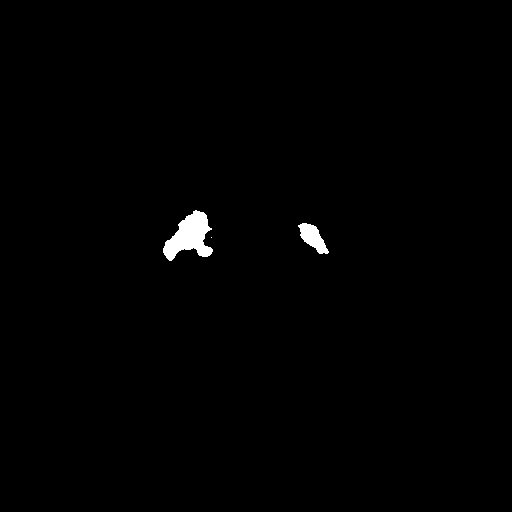}
		\includegraphics[width=0.13\textwidth,height = 0.13\textwidth]{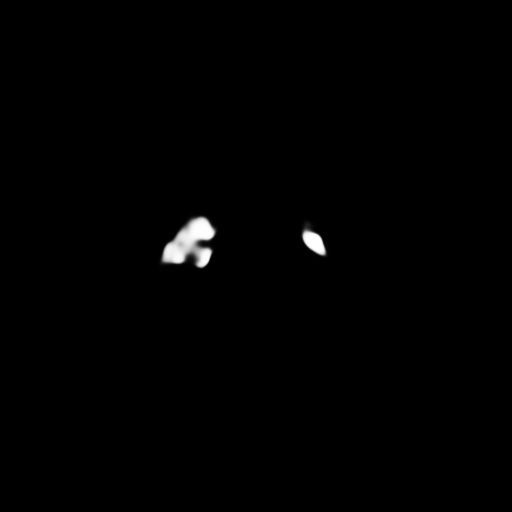}
		\includegraphics[width=0.13\textwidth,height = 0.13\textwidth]{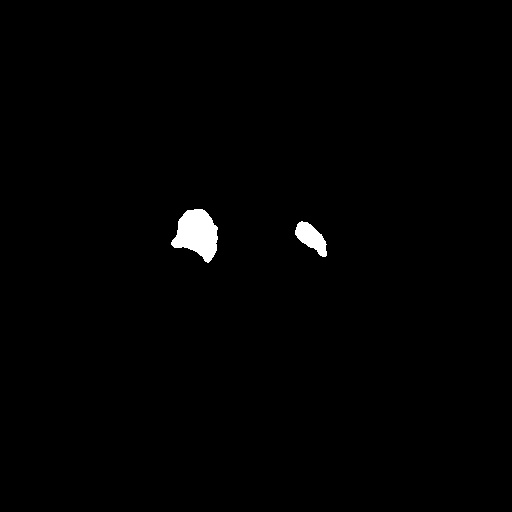}
		\includegraphics[width=0.13\textwidth,height = 0.13\textwidth]{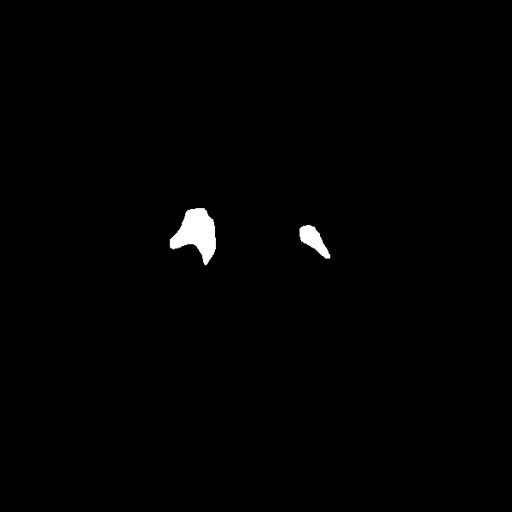}\\
		\vskip4pt
		\includegraphics[width=0.13\textwidth,height = 0.13\textwidth]{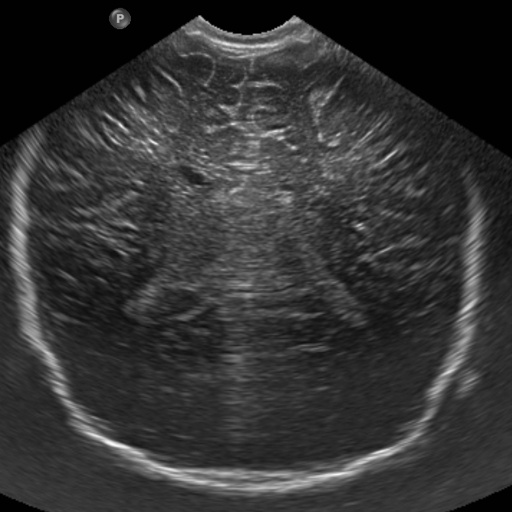}
		\includegraphics[width=0.13\textwidth,height = 0.13\textwidth]{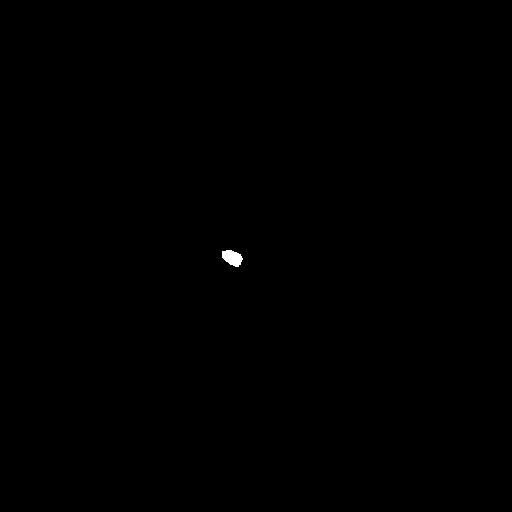}
		\includegraphics[width=0.13\textwidth,height = 0.13\textwidth]{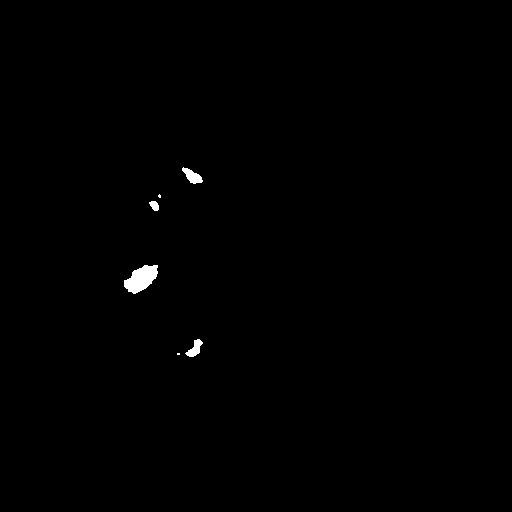}
		\includegraphics[width=0.13\textwidth,height = 0.13\textwidth]{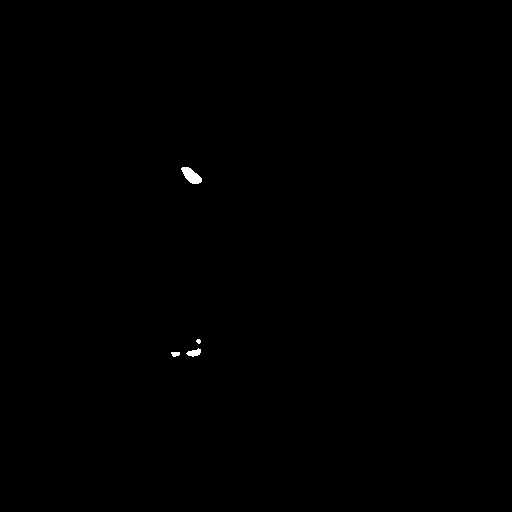}
		\includegraphics[width=0.13\textwidth,height = 0.13\textwidth]{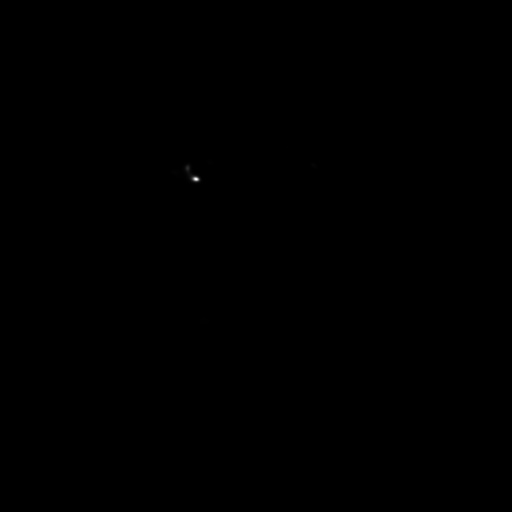}
		\includegraphics[width=0.13\textwidth,height = 0.13\textwidth]{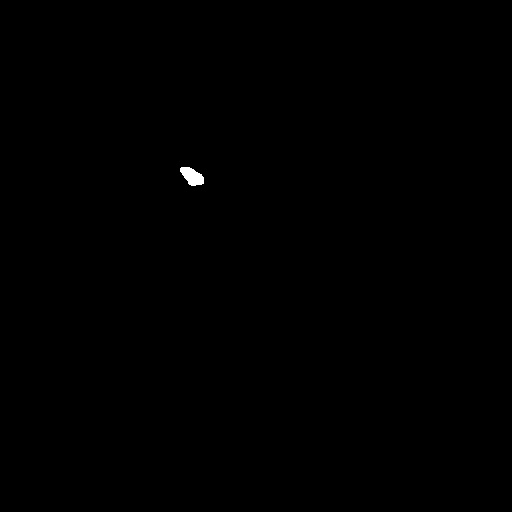}
		\includegraphics[width=0.13\textwidth,height = 0.13\textwidth]{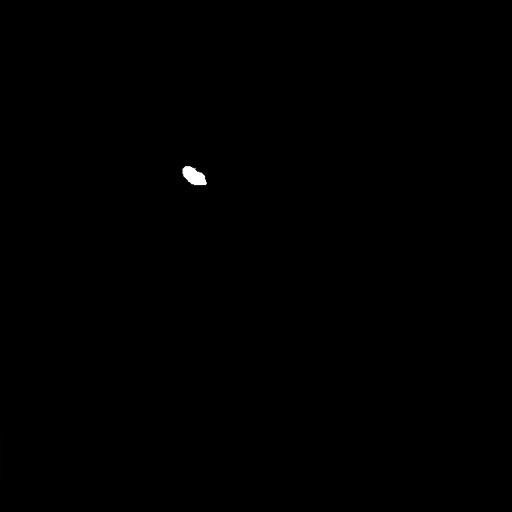}\\
		\vskip4pt
		\includegraphics[width=0.13\textwidth,height = 0.13\textwidth]{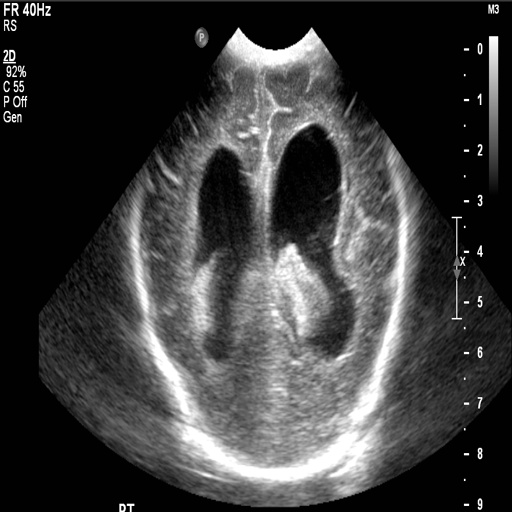}
		\includegraphics[width=0.13\textwidth,height = 0.13\textwidth]{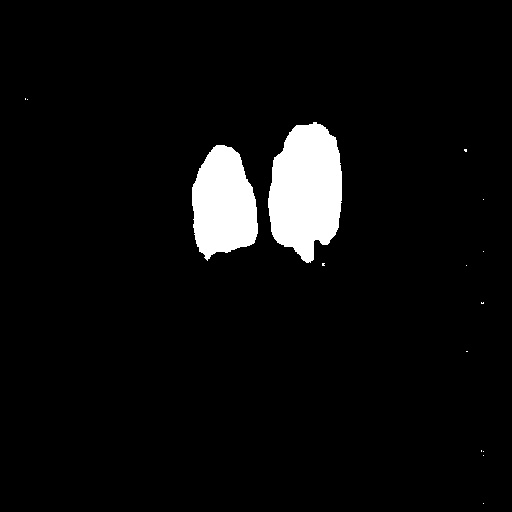}
		\includegraphics[width=0.13\textwidth,height = 0.13\textwidth]{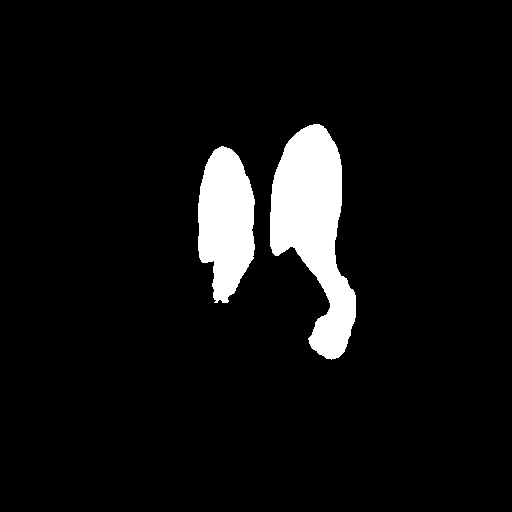}
		\includegraphics[width=0.13\textwidth,height = 0.13\textwidth]{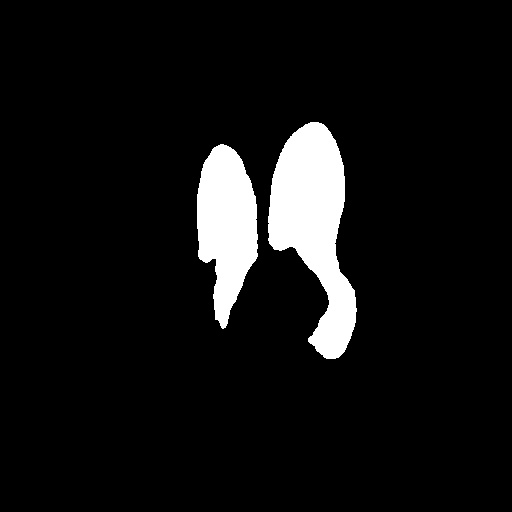}
		\includegraphics[width=0.13\textwidth,height = 0.13\textwidth]{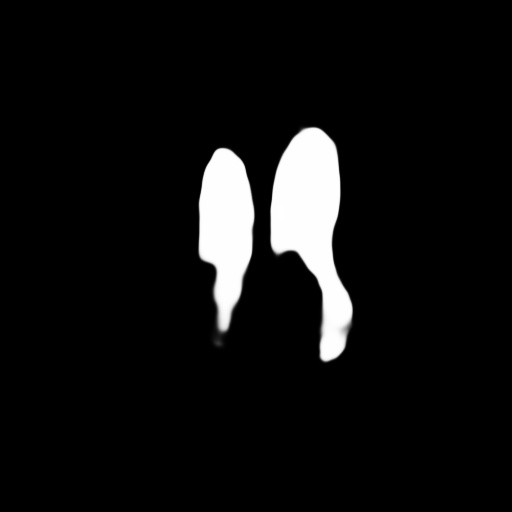}
		\includegraphics[width=0.13\textwidth,height = 0.13\textwidth]{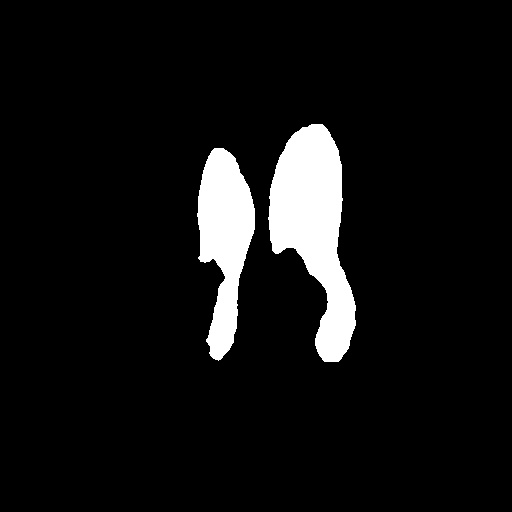}
		\includegraphics[width=0.13\textwidth,height = 0.13\textwidth]{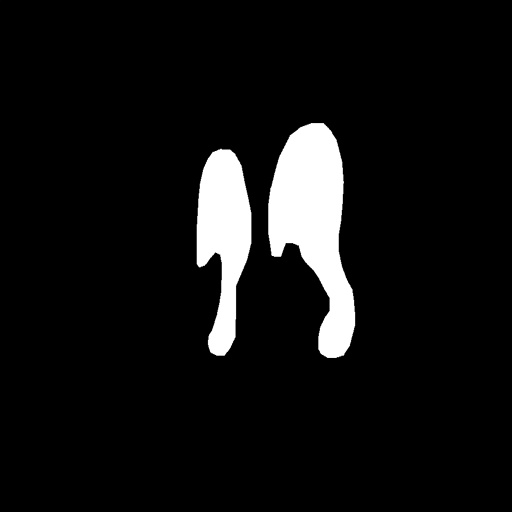}\\
		(a)\hskip60pt(b)\hskip60pt(c)\hskip60pt(d)\hskip60pt(e)\hskip60pt(f)\hskip60pt(g)
		\caption{Qualitative results on test images. (a) Input brain ultrasound image. (b) pix2pix \cite{isola2017image}.  (c) U-Net \cite{unet}. (d) UDe-Net \cite{unet,huang2017densely}. (e) Wang et.al \cite{wang2018automatic}. (f) CBAS (ours) (g) Ground-truth ventricle segmentation.}
		\label{Fig:exp2}
	\end{center}
\end{figure*}

\begin{table*}[htp!]
	\begin{center}
		\centering
		\caption{Comparison with pix2pix \cite{isola2017image}, U-Net\cite{unet}, UDe-Net\cite{unet,huang2017densely}, Wang et al.\cite{wang2018automatic}. Results shown correspond to mean values. %\hl{Can we include the computation times here as well? }
		}
		\resizebox{0.65\textwidth}{!}{
			\label{Comp}
			\begin{tabular}{c|c|c|c}
				\hline
				\hline
				Method       & DICE   & IoU(\%)  & Parameters \\ \hline 
				pix2pix\cite{isola2017image}        & 0.8584 $\pm$ 0.025 & 77.96 $\pm$ 0.031 & 11.1MB  \\ \hline
				U-Net\cite{unet}        & 0.8538 $\pm$ 0.024  & 77.90 $\pm$ 0.028 & 6.7MB  \\ \hline
				UDe-Net\cite{unet,huang2017densely}     & 0.8598 $\pm$ 0.017   & 78.09 $\pm$ 0.018  & 6.7MB \\ \hline
				Wang et.al\cite{wang2018automatic}   & 0.8725 $\pm$ 0.016   & 79.28 $\pm$ 0.014 & 24.9MB       \\ \hline
				CBAS & 0.8813 $\pm$ 0.008  & 80.25 $\pm$ 0.010 & 6.7MB \\ \hline 
				CBAS (with synthetic data generated using MSSA) & \textbf{0.8901 $\pm$ 0.063}  & \textbf{81.03 $\pm$ 0.061} & 6.7MB \\ \hline \hline
			\end{tabular}
		}
	\end{center}
\end{table*}

\section{Multi-Scale Self Attention (MSSA) guided Synthesis}
\label{sec:synthesis}
As the CBAS network is data-driven like most other deep learning methods, the performance of it is based on the size of training dataset. Collection of any medical image data and performing annotations of the same is a cumbersome and expensive process.  An approach to deal with this issue is to generate meaningful synthetic data which can be used to boost the segmentation performance.  To this end, we propose an image synthesis network that is trained to generate real-looking US images given the corresponding segmentation masks. Inspired from \cite{wang2018high}, we propose a multi-scale generator and discriminator networks to produce high-quality US images. Multi-scale networks have been used to generate stable high-resolution images \cite{isola2017image}.  However,  they still fail to capture  long-range dependencies in the US images. This makes the synthesized images look unrealistic with many artifacts near the edges of the anatomical structures.   To avoid this from happening, we propose a self-attention guided method where the self-attention module \cite{zhang2018self} is used to leverage the small-range capturing ability of the convolution blocks. The proposed network is called  Multi-Scale Self-Attention network (MSSA). 

\subsection{MSSA Network}

 Using the same notations as in Section \ref{sec:propose}, the problem statement can be viewed as an image translation task of synthesizing $\hat{x}$ from a given brain ventricle segmentation mask $s$. During training, a segmentation map $s$ such that $s\in\mathcal{S}$ is taken as input and its corresponding US scan $x$ such that $x\in\mathcal{B}$ is taken as the ground truth. The network we propose has a multi-scale generator architecture where the first part of the generator operates on the original scale of the segmentation mask ${s}$ and the second part of the generator operates on a down scaled (by 2) version of the segmentation mask ${s}_{\times 2}$. The proposed self-attention guided block operates on the down scaled version. Each self attention module \cite{vaswani2017attention}\cite{zhang2018self} has three 1$\times$1 convolution filters that are applied to the convolution feature maps. The output of each of the 1$\times$1 convolution layers can be represented as
 \begin{align}
\nonumber K(x) &= W_kx,\\
\nonumber Q(x) &= W_qx,\\
\nonumber V(x) &= W_vx,
 \end{align}
 where $W_k$,$W_q$ and $W_v$ are the 1$\times$1 convolution filters and $x$ is the convolutional feature maps. To get the self-attention feature maps, we perform dot product as follows
 $$\alpha_{i,j}=softmax(K(x_i)^TQ(x_j))$$
 $$o_j = \sum_{i=1}^N\alpha_{i,j}V(x_i)$$
 where $\alpha_{i,j}$ indicates the amount of attention the model gives while synthesizing the $j^{th}$ position from the $i^{th}$ location. The output self attention feature maps is the collection of the individual feature vectors $o_j$ where $j$ goes from $1$ to $N$. 
 
  The segmentation mask is first passed through a convolutional layer followed by an attention module which captures the dependencies of the image in its feature space. It is followed by a series of residual blocks \cite{he2016deep}. We use another self-attention module at the end of residual blocks to get the self-attention feature maps. These are concatenated with the feature maps that are generated from the generator at the original scale. The resulting concatenated feature maps are then further passed through the residual blocks before passing them through transpose convolution layers to get the US image. Owing to the high resolution of the synthesized image, we use a two scale discriminator that works on the original as well as the down scaled (by 2) version of the real and synthesized image. The discriminator architecture across both scales are patch based fully convolutional networks \cite{long2015fully}. It should be noted that more scales can be added to the proposed network if the computation time is not of a concern. The Generator architecture has the following sequence of blocks:
 
 Half Scale Part:
 \\
  ConvBlock1(1,64)
  \\
  ConvBlock2(128)-ConvBlock2(256)-
  \\
  ConvBlock2(512)-ConvBlock2(1024)-
  SelfAttentionBlock,
  \\
  ResBlock(1024)$\times$ 9, 
  \\
  ConvBlock3(512)-ConvBlock3(256)-
  \\
  ConvBlock3(128)-ConvBlock3(64)-
  \\ConvBlock1(1,1)-SelfAttentionBlock.
  \\
  
  Full-Scale Part:
 \\
  ConvBlock1(1,32)-ConvBlock2(64)-
  \\(Output of this is added with the self attention maps from the Half Scale part.)
  \\ResBlock(64)$\times$ 3,
  \\ConvBlock3(32)-ConvBlock1(1,1). 
  
  The discriminator architecture has the following sequence of blocks:\\
  ConvBlock(64),ConvBlock(128),
  \\
  ConvBlock(256),ConvBlock(512).
  
  The details about the layers in each of the above blocks is explained in the appendix. The overall network architecture is illustrated in Fig \ref{Fig:MSSA}. 

\subsection{Loss for MSSA Network}
Let $G$ denote the generator network and $D_1$, $D_2$ denote the discriminator networks. Our objective function to train the overall network is as follows
\begin{equation}\label{Eq:eq6}
\min_G ((\max_{D_1,D_2} \sum_{k=1,2}\mathcal{L}_{GAN}(G,D_k)) + \lambda_{1}\sum_{k=1,2}\mathcal{L}_{FM}(G,\mathcal{D}_k)),
\end{equation}
where 
\begin{align}
\nonumber \mathcal{L_{GAN}}(G,D_k)&=\E_{(x,s)}\log D_k(x,s)\\ &+ \E_x[(\log(1-D_k(x,G(x)))],
\end{align}
and 
$$\mathcal{L_{FM}}(G,D_k)=\E_{(x,s)}\sum_{i=1}^{T}\dfrac{1}{N_i}\|[\log D_k^i(x,s)-D_k^i(x,G(x))]\|_{2}^{2}$$
are the two loss functions in the overall objective function. It can be noted that $x$ is the US scan that is to be synthesized and $s$ is the input segmentation mask. $\mathcal{L}_{GAN}$ is the standard GAN loss which is the sum of expectation over the discriminator's estimate of how much probability that the data instance is real/fake depending on whether it is a real data instance or if it is synthesized from the generator. $\mathcal{L}_{FM}$ is the feature matching loss which is  a perceptual loss \cite{johnson2016perceptual} calculated across different layers in the discriminator. $\lambda_{1}$ is the factor which controls the amount of feature matching loss that is to affect objective function. $N_i$ denotes the number of elements in the $i^{th}$ layer and $T$ denotes the total number of layers in the network.

\section{Experiments and Results}
\label{exp_res}
In  this  section,  we  present  details  of  the  experiments  and quality  measures  used  to  evaluate  the  proposed synthesis and segmentation methods.  We  also  discuss  the  dataset  and  training  details followed by comparison of the proposed methods against a set of baseline methods and recent state-of-the-art approaches.
\subsection{Dataset}
\begin{figure}[htp!]
		\centering
		\includegraphics[width=0.2\textwidth,height=0.2\textwidth]{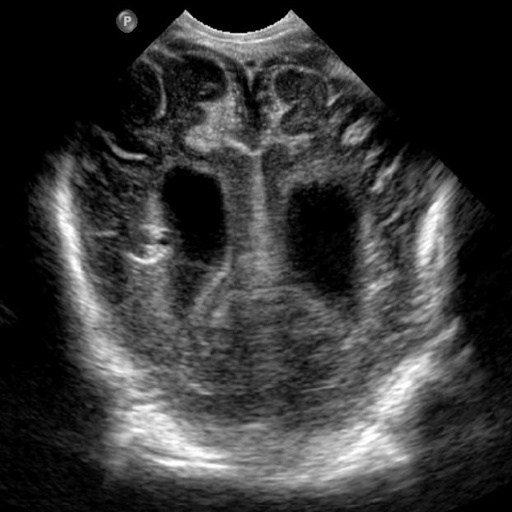}
		\hskip3pt
		\includegraphics[width=0.2\textwidth,height=0.2\textwidth]{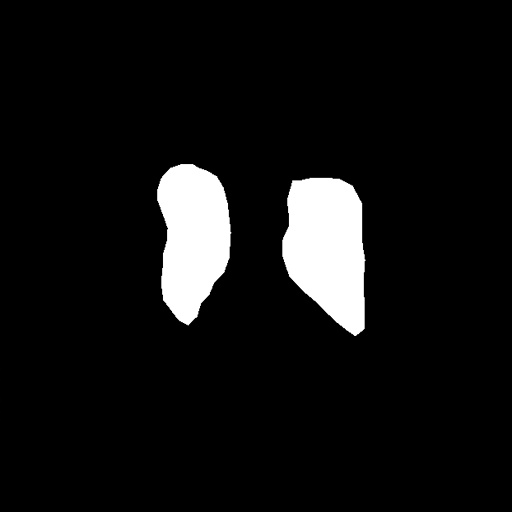}
		\vskip4pt
		\includegraphics[width=0.2\textwidth,height=0.2\textwidth]{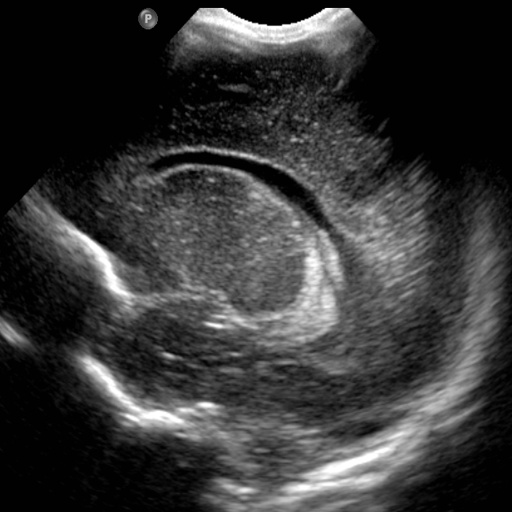}
		\hskip3pt
		\includegraphics[width=0.2\textwidth,height=0.2\textwidth]{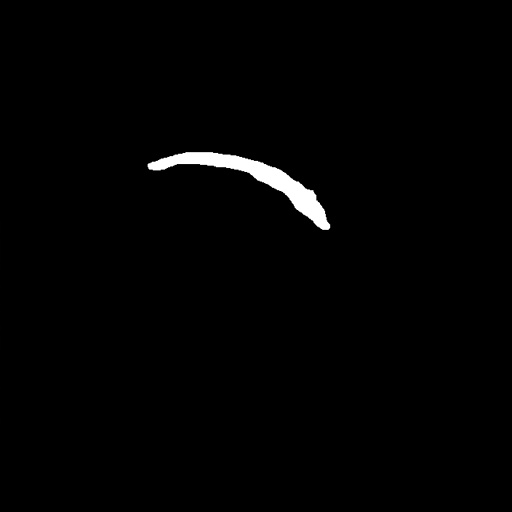}\\
		(a)\hskip100pt(b)
		\caption{Sample brain ultrasound images from the dataset. (a) brain US images.  (b) the corresponding segmentation masks.}
	\label{fig:sampledata}
\end{figure}

After obtaining institutional review board (IRB) approval,  retrospective brain US scans were collected. 
A total of 1629 in vivo B-mode US images were obtained from 20 different subjects (age\textless1 years old) who were treated between 2010 and 2016. The dataset contained subjects with IVH and without (healthy subjects but in risk of developing IVH). The US scans were collected using a Philips US machine with a C8-5 broadband curved array transducer using coronal and sagittal scan planes. For every collected image ventricles and septum pellecudi are manually segmented by an expert ultrasonographer. We split these images randomly into 1300 Training images and 329 Testing images for experiments. Note that these images are of size $512 \times 512$. During the random split of the dataset the training and testing data did not include the same patient scans.  Sample images and the corresponding segmentation masks from this dataset are shown in Fig.~\ref{fig:sampledata}.

We evaluate the performance of both our segmentation and synthesis methods with recent methods on the randomly selected 329 test images. We compare the performance of our synthesis method against, pix2pix \cite{isola2017image}, U-Net \cite{unet}, UDe-Net \cite{huang2017densely}, and  Wang et al. \cite{wang2018automatic}. We conduct these experiments three times and average out the obtained results. We use DICE coefficient and Intersection over Union (IoU) to measure the performance of different segmentation networks. For the synthesis network, we compare our method with pix2pix \cite{isola2017image} and pix2pixHD \cite{wang2018high}. We calculate Structural Similarity Index (SSIM) between the synthesized images and the real images as well as the DICE accuracy of CBAS when trained with the synthetic data generated from each of the compared methods to validate the performance of our method apart from the qualitative results.

\subsection{Training Details}
CBAS is trained using $\mathcal{L}_{final}$ with the Adam optimizer \cite{kingma2014adam} and batch size of 1. The learning rate is set equal to 0.0002 and annealed by $5\%$ for every 10 epochs. $\lambda$ is set equal to $0.1$ for initial epochs, but when the mean of all values in the confidence maps $c_{\times 1}, c_{\times 2}$ is greater than $0.75$ then $\lambda$ is set equal to $0.01$. CBAS is trained for 100 epochs. We perform data augmentation using horizontal, vertical flips and random crops to extend the training images to 6500 images. We resize the images to $640 \times 640$ and crop $512 \times 512$ size patches to obtain random crop images.

MSSA is trained  using a learning rate of 0.0002 with the Adam optimizer \cite{kingma2014adam} and batch size of 1. The half-scale self-attention guided part of the generator is trained separately for the first 200 epochs.  Then, the full scale part of the network is trained along with this for the next 300 epochs.  $\lambda_{1}$ in Eq.~\eqref{Eq:eq6} is set equatl to 0.1.

\begin{figure*}[htp!]
	\begin{center}
		\centering
		\includegraphics[width=0.18\textwidth,height = 0.18\textwidth]{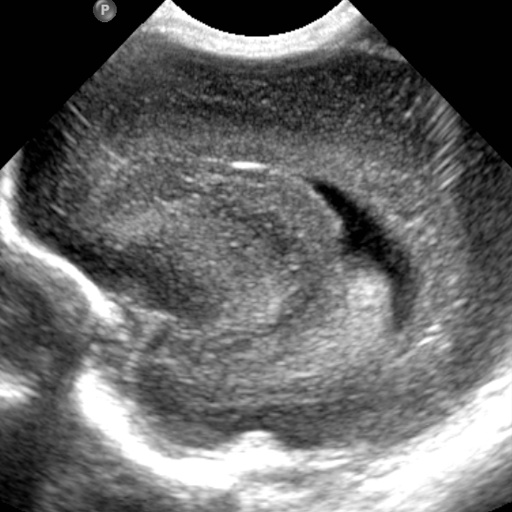}
		\includegraphics[width=0.18\textwidth,height = 0.18\textwidth]{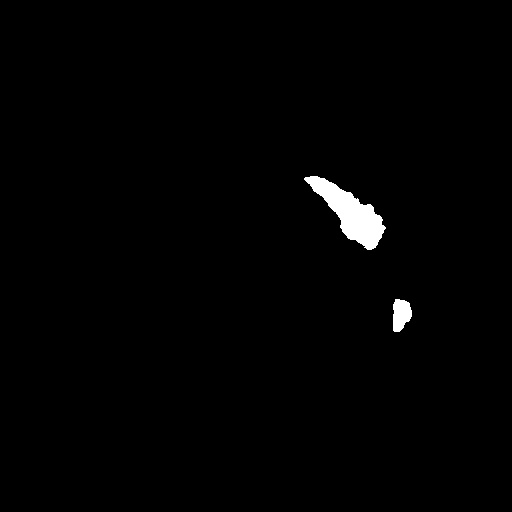}
		\includegraphics[width=0.18\textwidth,height = 0.18\textwidth]{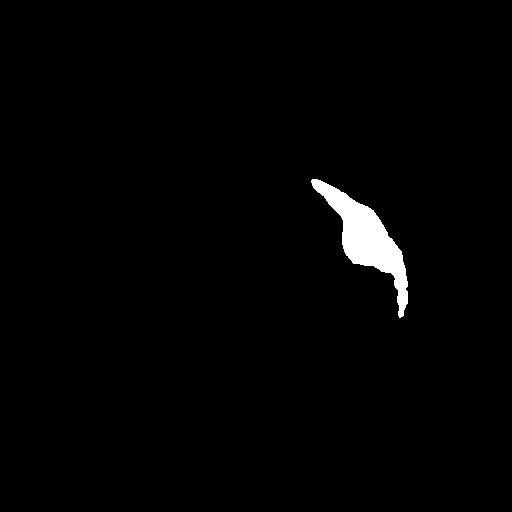}
		\includegraphics[width=0.18\textwidth,height = 0.18\textwidth]{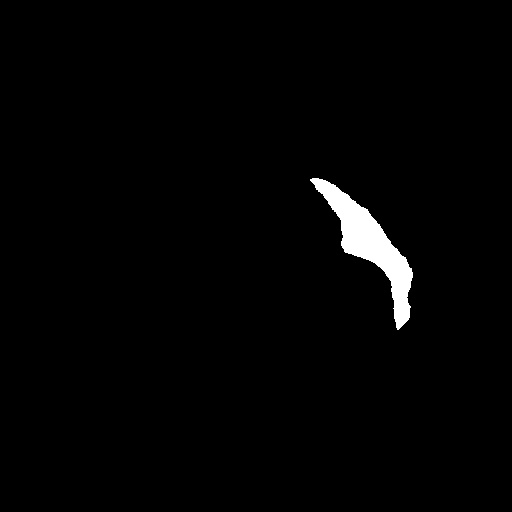}
		\includegraphics[width=0.18\textwidth,height = 0.18\textwidth]{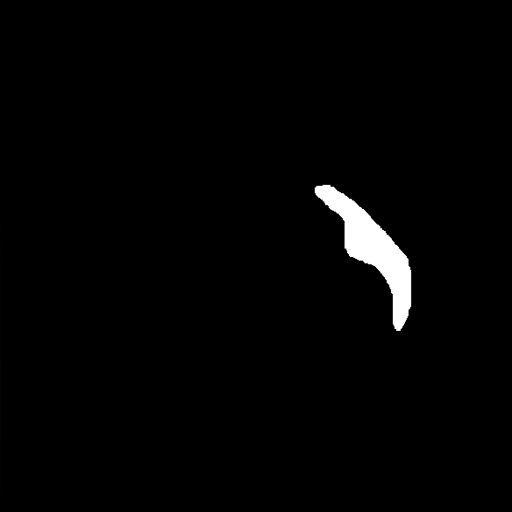}\\
		\vskip4pt
		\includegraphics[width=0.18\textwidth,height = 0.18\textwidth]{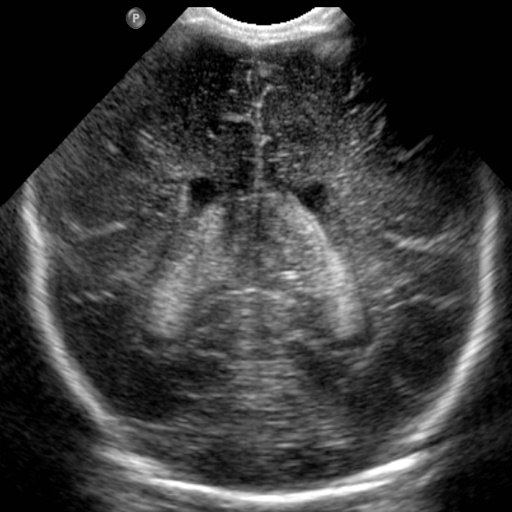}
		\includegraphics[width=0.18\textwidth,height = 0.18\textwidth]{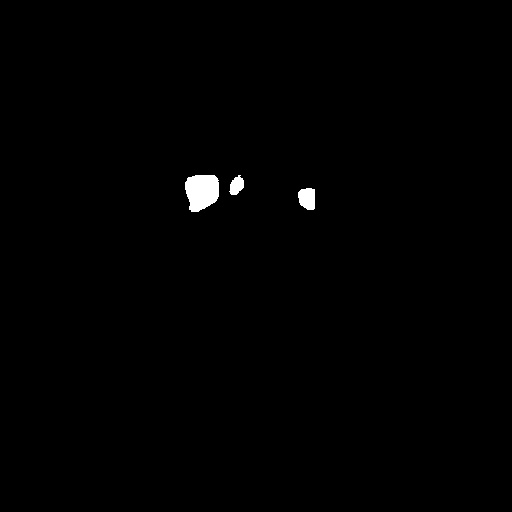}
		\includegraphics[width=0.18\textwidth,height = 0.18\textwidth]{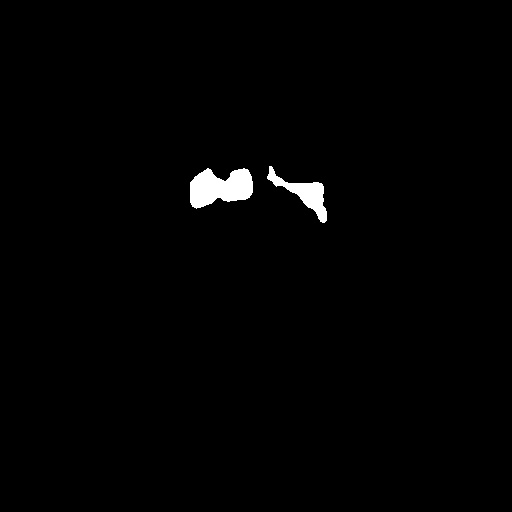}
		\includegraphics[width=0.18\textwidth,height = 0.18\textwidth]{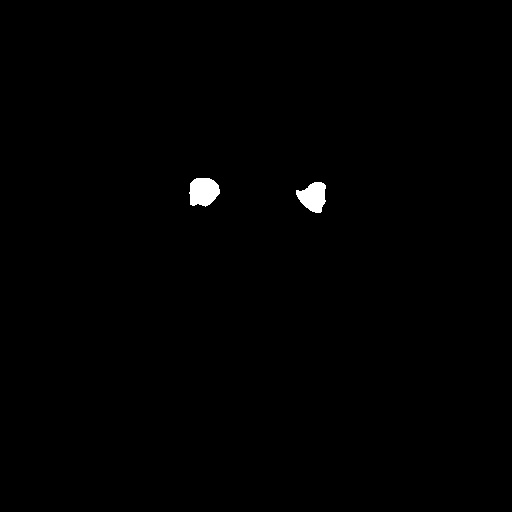}
		\includegraphics[width=0.18\textwidth,height = 0.18\textwidth]{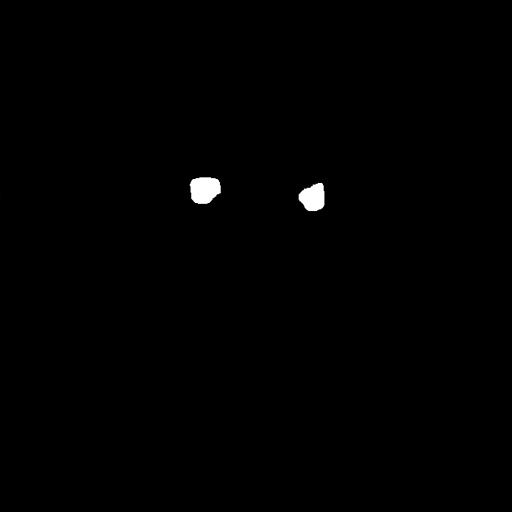}\\
		\vskip4pt
		\includegraphics[width=0.18\textwidth,height = 0.18\textwidth]{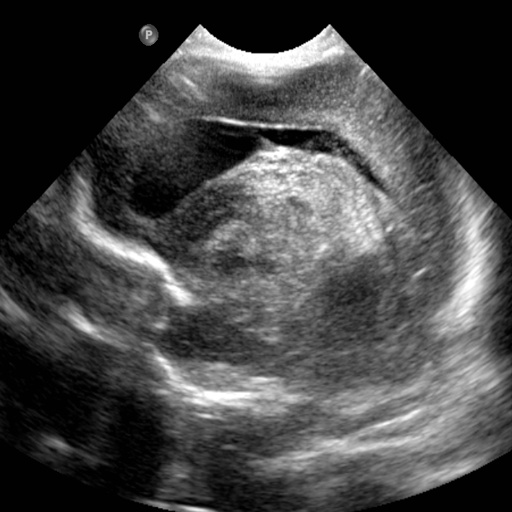}
		\includegraphics[width=0.18\textwidth,height = 0.18\textwidth]{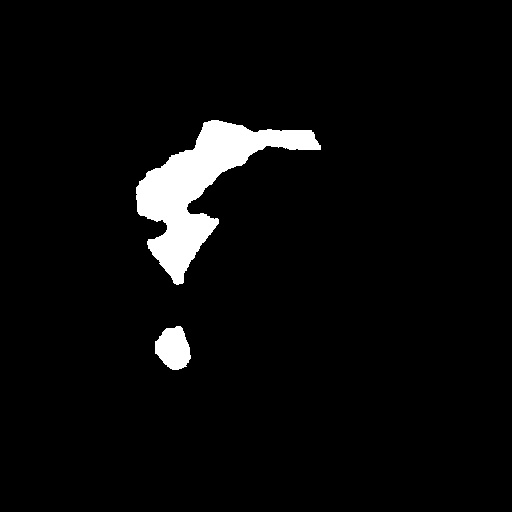}
		\includegraphics[width=0.18\textwidth,height = 0.18\textwidth]{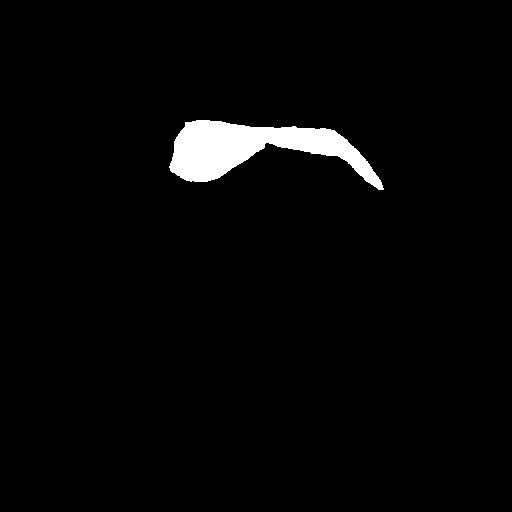}
		\includegraphics[width=0.18\textwidth,height = 0.18\textwidth]{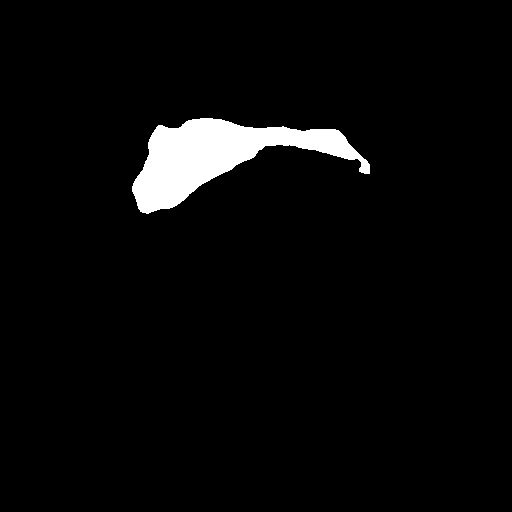}
		\includegraphics[width=0.18\textwidth,height = 0.18\textwidth]{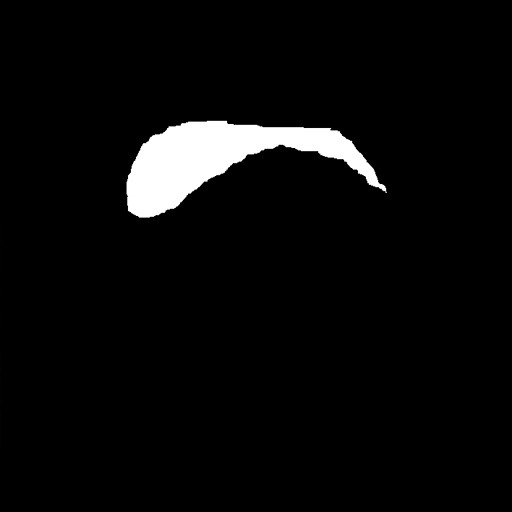}\\
		(a)\hskip90pt(b)\hskip90pt(c)\hskip90pt(d)\hskip90pt(e)\\
		\caption{Qualitative results on test images for ablation study. (a) Input brain ultrasound image, (b) BN, (c) BN w/ SB, (d) CBAS (ours), (e) Ground-truth ventricle segmentation. Brain ventricular segmentations from BN and BN w/ SB have over segmentation for small size ventricles, and incorrect segmentation at the edges of large size ventricles. CBAS trained with $\mathcal{L}_{final}$ produced best results with accurate segmentation for both small and large sized ventricles.}
		\label{Fig:exp3}
	\end{center}
\end{figure*}

\begin{table*}[htp!]
	\centering
	\begin{center}
		\caption{Quantitative results corresponding to ablation study. Results shown correspond to mean values.}
		\resizebox{0.6\textwidth}{!}{
			\label{ABL}
			\begin{tabular}{c|c|c|c|c}
				\hline \hline
				Method  & Loss & DICE  & IoU(\%) & \textit{p}-value \\ \hline
				BN     & CE & 0.8538   &  77.90  &  6.48 $\times 10^{-3}$  \\ \hline
				BN w/ SB & CE & 0.8673  & 78.48 &  3.34 $\times 10^{-2}$    \\ \hline
				BN w/ SB and CB & CE & 0.8664   & 78.56   & 4.74 $\times 10^{-2}$    \\ \hline
				CBAS   & $\mathcal{L}_{final}$ & 0.8813  & 80.25 & -- \\ \hline 
				CBAS (with synthetic data)  & $\mathcal{L}_{final}$ & \textbf{0.8901} & \textbf{81.03} & -- \\ \hline \hline
			\end{tabular}
		}
	\end{center}
\end{table*}

\begin{figure*}[htp!]
	\begin{center}
		\centering
		\includegraphics[width=0.2\textwidth,height = 0.15\textwidth]{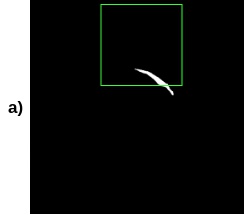}
		\includegraphics[width=0.18\textwidth,height = 0.15\textwidth]{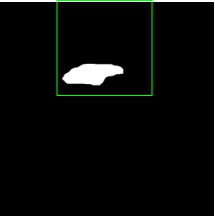}
		\includegraphics[width=0.18\textwidth,height = 0.15\textwidth]{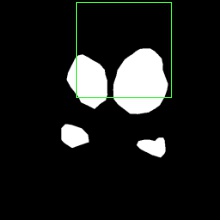}
		\includegraphics[width=0.18\textwidth,height = 0.15\textwidth]{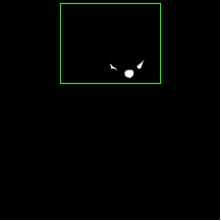}
		\includegraphics[width=0.18\textwidth,height = 0.15\textwidth]{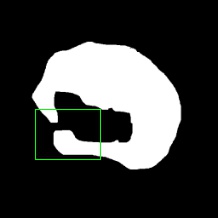}\\
		\vskip4pt
		\includegraphics[width=0.2\textwidth,height = 0.15\textwidth]{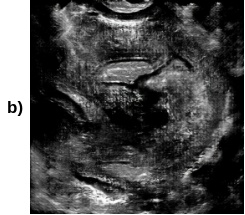}
		\includegraphics[width=0.18\textwidth,height = 0.15\textwidth]{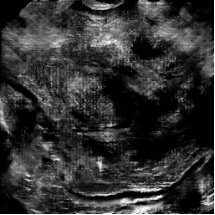}
		\includegraphics[width=0.18\textwidth,height = 0.15\textwidth]{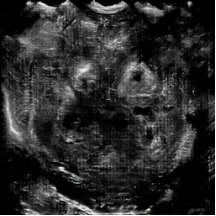}
		\includegraphics[width=0.18\textwidth,height = 0.15\textwidth]{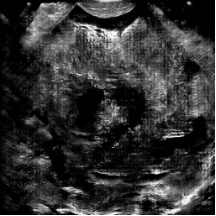}
		\includegraphics[width=0.18\textwidth,height = 0.15\textwidth]{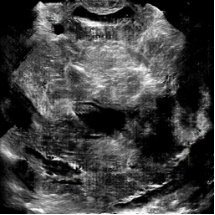}\\
		\vskip4pt
		\includegraphics[width=0.2\textwidth,height = 0.15\textwidth]{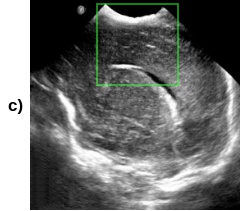}
		\includegraphics[width=0.18\textwidth,height = 0.15\textwidth]{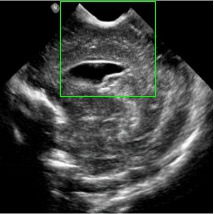}
		\includegraphics[width=0.18\textwidth,height = 0.15\textwidth]{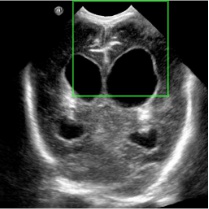}
		\includegraphics[width=0.18\textwidth,height = 0.15\textwidth]{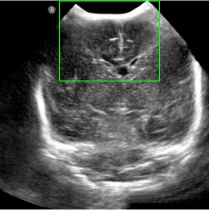}
		\includegraphics[width=0.18\textwidth,height = 0.15\textwidth]{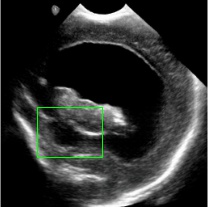}\\
		\vskip4pt
		\includegraphics[width=0.2\textwidth,height = 0.15\textwidth]{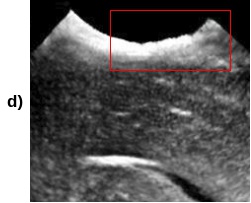}
		\includegraphics[width=0.18\textwidth,height = 0.15\textwidth]{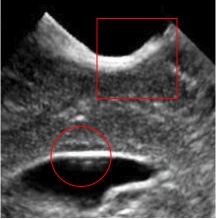}
		\includegraphics[width=0.18\textwidth,height = 0.15\textwidth]{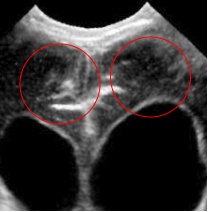}
		\includegraphics[width=0.18\textwidth,height = 0.15\textwidth]{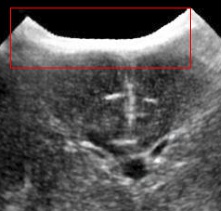}
		\includegraphics[width=0.18\textwidth,height = 0.15\textwidth]{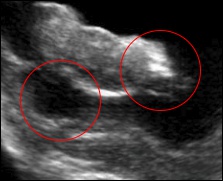}\\
		\vskip4pt
		\includegraphics[width=0.2\textwidth,height = 0.15\textwidth]{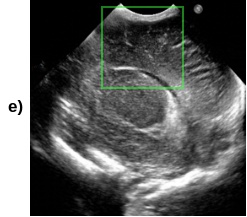}
		\includegraphics[width=0.18\textwidth,height = 0.15\textwidth]{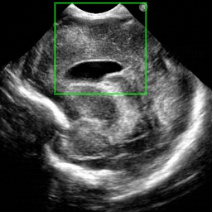}
		\includegraphics[width=0.18\textwidth,height = 0.15\textwidth]{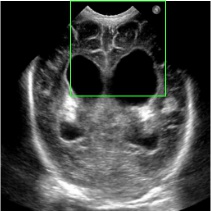}
		\includegraphics[width=0.18\textwidth,height = 0.15\textwidth]{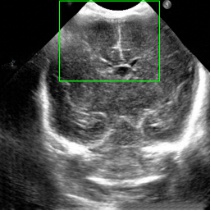}
		\includegraphics[width=0.18\textwidth,height = 0.15\textwidth]{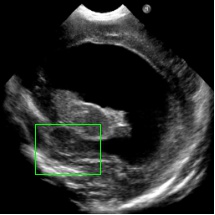}\\
		\vskip4pt
		\includegraphics[width=0.2\textwidth,height = 0.15\textwidth]{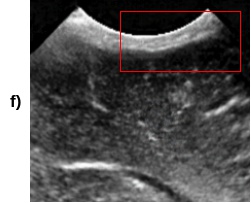}
		\includegraphics[width=0.18\textwidth,height = 0.15\textwidth]{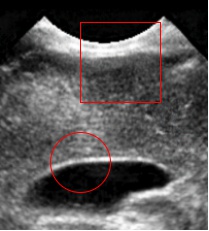}
		\includegraphics[width=0.18\textwidth,height = 0.15\textwidth]{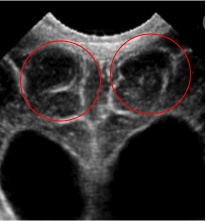}
		\includegraphics[width=0.18\textwidth,height = 0.15\textwidth]{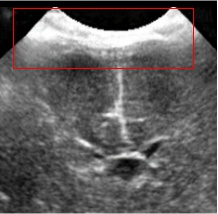}
		\includegraphics[width=0.18\textwidth,height = 0.15\textwidth]{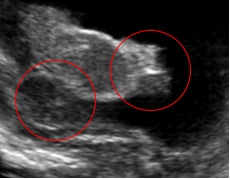}\\
		\vskip4pt
		\includegraphics[width=0.2\textwidth,height = 0.15\textwidth]{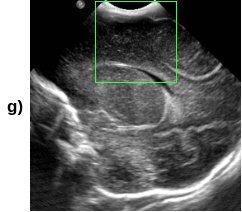}
		\includegraphics[width=0.18\textwidth,height = 0.15\textwidth]{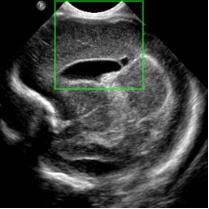}
		\includegraphics[width=0.18\textwidth,height = 0.15\textwidth]{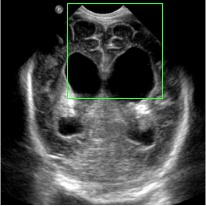}
		\includegraphics[width=0.18\textwidth,height = 0.15\textwidth]{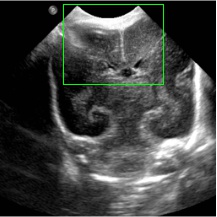}
		\includegraphics[width=0.18\textwidth,height = 0.15\textwidth]{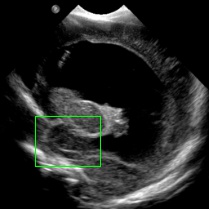}\\
		\vskip4pt
		\includegraphics[width=0.2\textwidth,height = 0.15\textwidth]{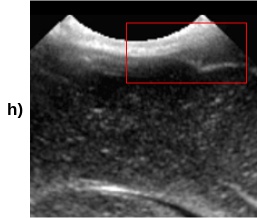}
		\includegraphics[width=0.18\textwidth,height = 0.15\textwidth]{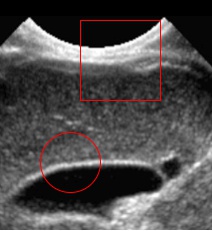}
		\includegraphics[width=0.18\textwidth,height = 0.15\textwidth]{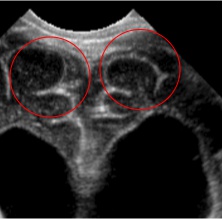}
		\includegraphics[width=0.18\textwidth,height = 0.15\textwidth]{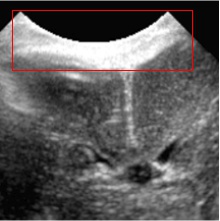}
		\includegraphics[width=0.18\textwidth,height = 0.15\textwidth]{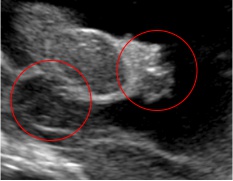}\\
		
		\caption{Qualitative results on test images for the synthesis task. (a) Input segmentation mask, (b) Synthesized image using pix2pix \cite{isola2017image}, (c) Synthesized image using pix2pixHD \cite{wang2018high}, (e) Synthesized image using MSSA (ours), (g) Real B-Mode Ultrasound image for the input segmentation mask in (a). Images shown in (d),(f) and (h) are the zoomed in parts inside the green box as shown in (c),(e) and (f) respectively. The red boxes in (d),(f) and (h) denote the specific structures that show how our method is closer to the real image than pix2pixHD \cite{wang2018high}.}
		\label{Fig:synallex}
		\label{Fig:exp4}
	\end{center}
\end{figure*}

\begin{figure*}[htp!]
	\begin{center}
		\centering
		\includegraphics[width=0.15\textwidth,height = 0.12\textwidth]{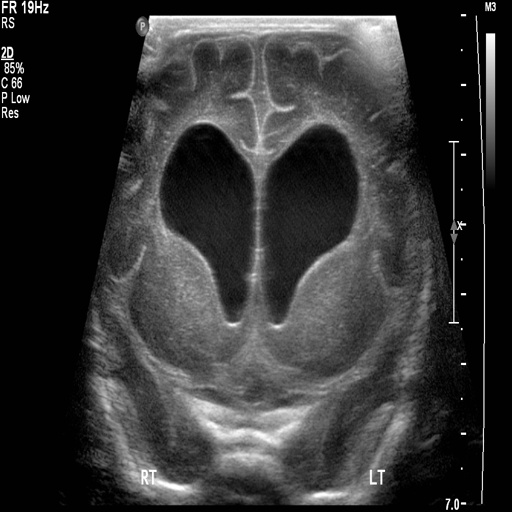}
		\includegraphics[width=0.15\textwidth,height = 0.12\textwidth]{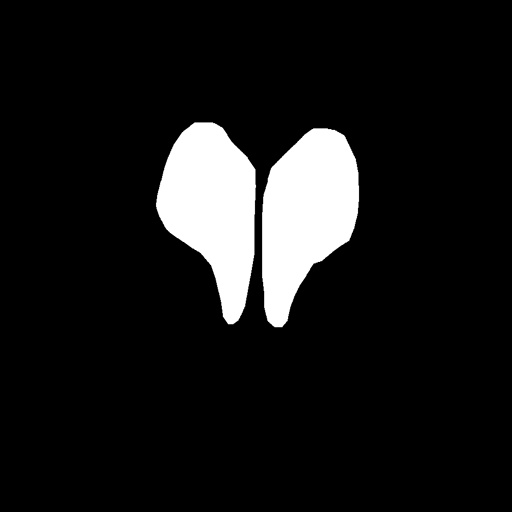}
		\includegraphics[width=0.15\textwidth,height = 0.12\textwidth]{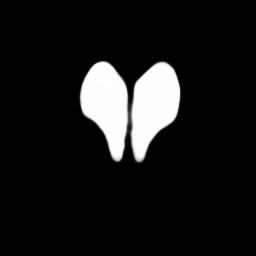}
		\includegraphics[width=0.15\textwidth,height = 0.12\textwidth]{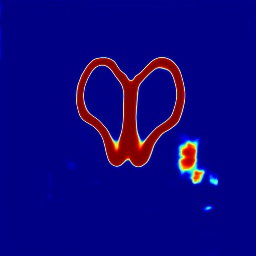}
		\includegraphics[width=0.15\textwidth,height = 0.12\textwidth]{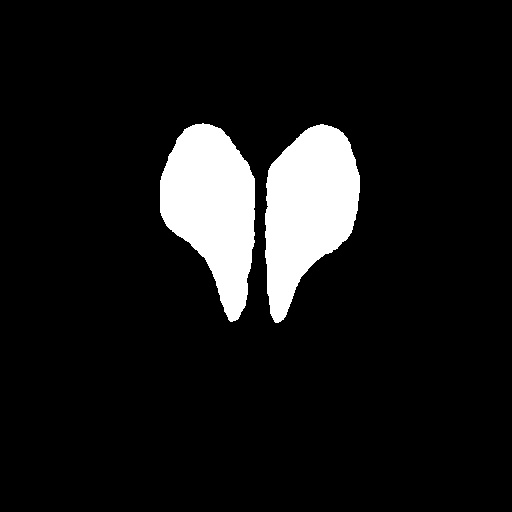}
		\includegraphics[width=0.15\textwidth,height = 0.12\textwidth]{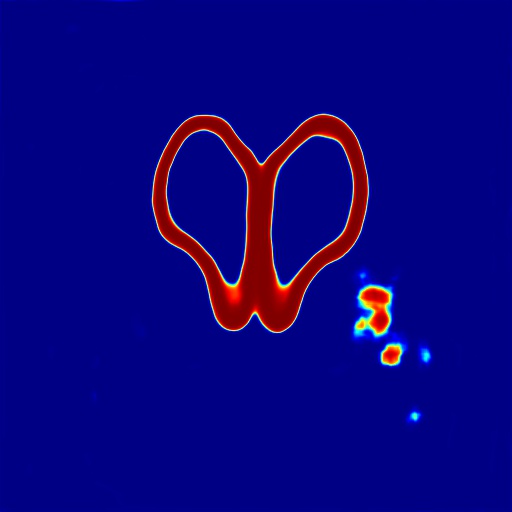}\\
		\vskip4pt
		\includegraphics[width=0.15\textwidth,height = 0.12\textwidth]{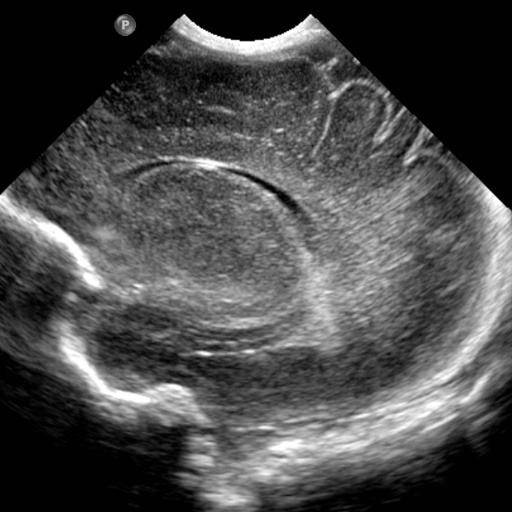}
		\includegraphics[width=0.15\textwidth,height = 0.12\textwidth]{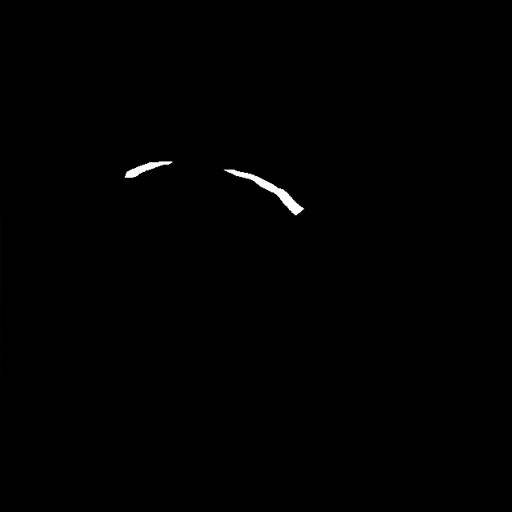}
		\includegraphics[width=0.15\textwidth,height = 0.12\textwidth]{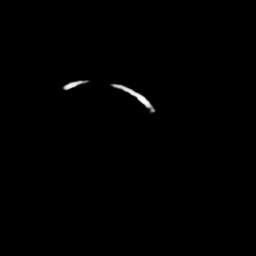}
		\includegraphics[width=0.15\textwidth,height = 0.12\textwidth]{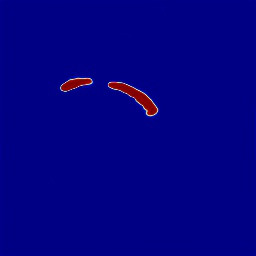}
		\includegraphics[width=0.15\textwidth,height = 0.12\textwidth]{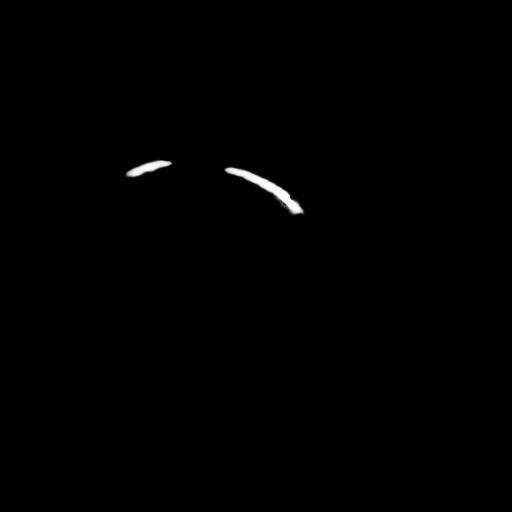}
		\includegraphics[width=0.15\textwidth,height = 0.12\textwidth]{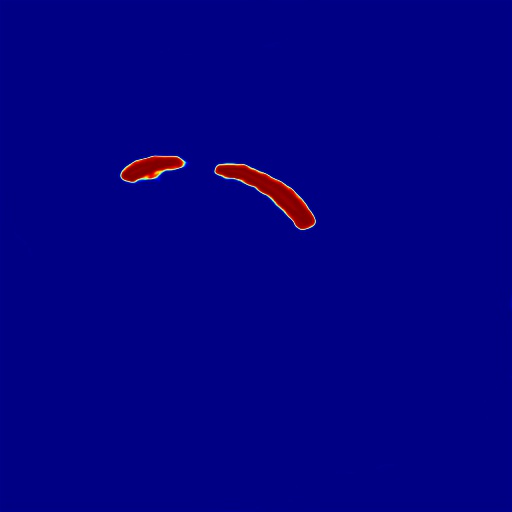}\\
		\vskip4pt
		\includegraphics[width=0.15\textwidth,height = 0.12\textwidth]{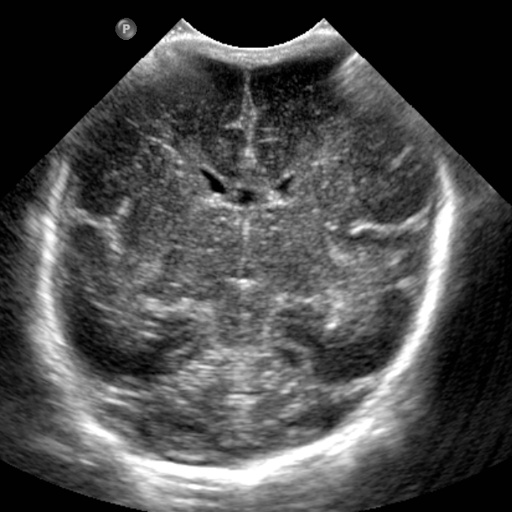}
		\includegraphics[width=0.15\textwidth,height = 0.12\textwidth]{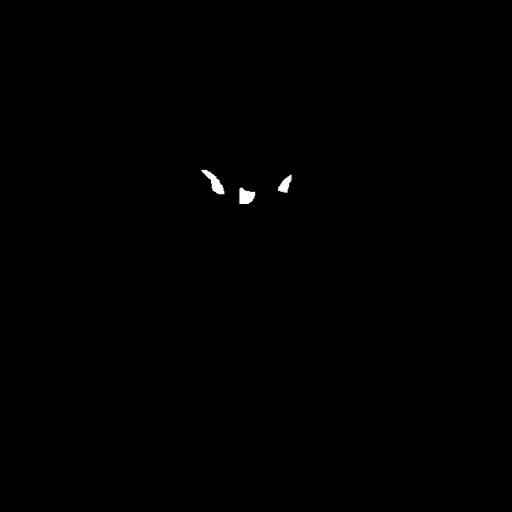}
		\includegraphics[width=0.15\textwidth,height = 0.12\textwidth]{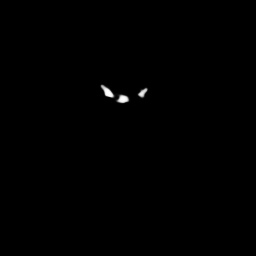}
		\includegraphics[width=0.15\textwidth,height = 0.12\textwidth]{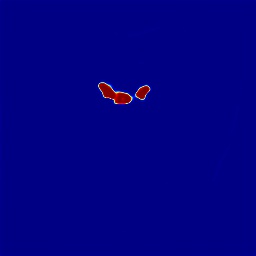}
		\includegraphics[width=0.15\textwidth,height = 0.12\textwidth]{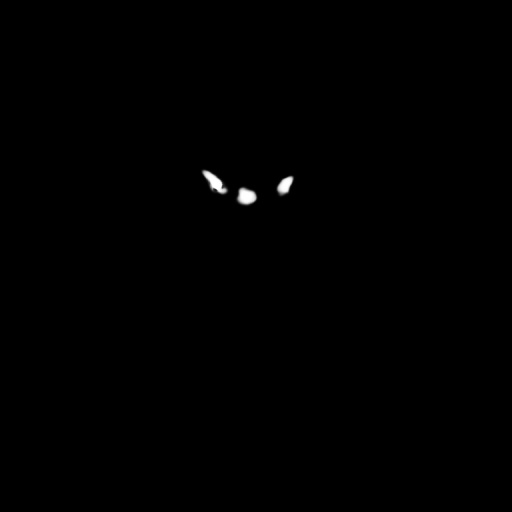}
		\includegraphics[width=0.15\textwidth,height = 0.12\textwidth]{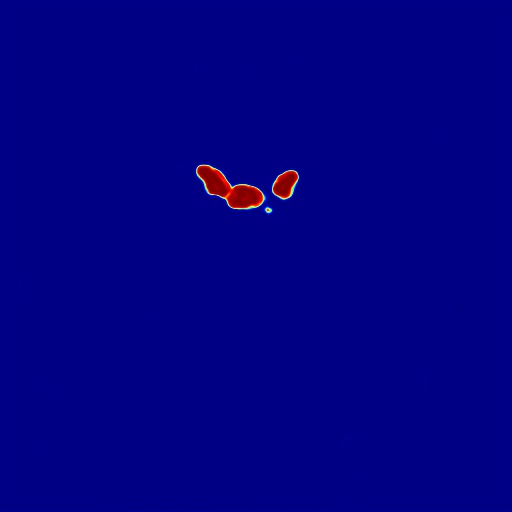}\\
		(a)\hskip70pt(b)\hskip70pt(c)\hskip70pt(d)\hskip70pt(e)\hskip70pt(f)\\
		\caption{Confidence maps visualization for a test image. (a) Input brain ultrasound image. (b) Ground-truth ventricle segmentation. (c), (e) are the estimated vetricle segmentations at different scales $\hat{s}_{\times 2}$, and $\hat{s}_{\times 1}$, respectively. (d),(f) are the corresponding confidence maps $c_{\times 2}$, and $c_{\times 1}$, respectively. Note that in the confidence maps blue means 1 and red means 0}
		\label{Fig:exp4}
	\end{center}
\end{figure*}

\subsection{Qualitative Performance}
Fig.~\ref{Fig:exp2} shows the qualitative performance of different segmentation methods on the test images. We can clearly see that pix2pix \cite{isola2017image}, U-Net \cite{unet}, UDe-Net \cite{unet,huang2017densely}, and Wang et al. \cite{wang2018automatic}, misclassified normal regions as the brain ventricular regions. For example, from the second column of Fig.~\ref{Fig:exp2}, we can clearly observe under segmentation of brain ventricles regions in the outputs produced using pix2pix \cite{isola2017image}. Brain ventricle segmentations obtained using U-Net \cite{unet}, and UDe-Net \cite{huang2017densely} also contain under segmentation for large size ventricles (in the fourth row) and over segmentation for small size ventricles as shown in the third and the fourth columns of  Fig.~\ref{Fig:exp2}. Wang et al. \cite{wang2018automatic} produce brain  segmentations which contain inaccurate edges for large ventricles and under segmentation for small size ventricles. On the other hand, the estimated shape of the brain ventricular regions by those methods are slightly off when compared to the original shape. Visually we can see that CBAS produces more accurate brain ventricular regions, and does not miss-classify the normal regions as brain ventricular regions.

Fig.~\ref{Fig:synallex} shows the qualitative performance of different synthesis methods. We observe that pix2pix \cite{isola2017image} is very unstable at generating high-resolution images and performs very poorly in almost every case. The pix2pixHD method \cite{wang2018high} synthesizes high-resolution images but fails to synthesize realistic looking images. In the second row, it can be observed that pix2pixHD does not properly capture the features of the US image near the edges. Similarly, as can be seen from the third, fourth and fifth rows, the structures inside the US image are not captured by pix2pixHD. Our proposed method captures all these structures that are missed by pix2pixHD which can be seen in the illustration.      

\subsection{Quantitative Performance}
Table.~\ref{Comp} shows the quantitative performance of our proposed segmentation method and the other investigated methods. As it can be seen from this table, our method clearly outperforms these recent segmentation methods (p\textless0.05 for paired t-test with 5\% significance). The paired t-test value using the DICE scores between CBAS and Wang et.al\cite{wang2018automatic} (second best method), resulted in an average $p$ value of $3.63 \times 10^{-9}$. Note that, our method has very less number of parameters in the network as compared to Wang et al. \cite{wang2018automatic}. Time taken by our method to process an image of $512 \times 512$ is about 0.01 seconds compared to 0.02 seconds for \cite{wang2018automatic}. This presents a 50\% improvement in computation time. 

Table ~\ref{Comp_syn} shows the quantitative performance of our proposed synthesis method and the other recent methods. The DICE accuracy is calculated by training our proposed segmentation method (CBAS) on equal proportion of real and synthetic images, where the synthetic images are generated by the methods we compare. We use a total of 2600 (1300 real and 1300 synthetic) images to train our CBAS network. SSIM measure is found with the 329 segmentation masks and real images that were left out during training the synthesis network. The synthesis method being compared is used to feed in the segmentation map and the resultant synthesized images are compared with the real images for SSIM. From Table~~\ref{Comp_syn}, it can be observed that the segmentation network performs the best when the synthesized images added are generated using our proposed method. Furthermore, SSIM results in Table~~\ref{Comp_syn} indicate that our method generates better quality images than the other methods. It can be noted that the addition of self attention to the base network \cite{wang2018high} improved the qualitative and quantitative results as seen in Table \ref{Comp_syn} and Fig \ref{Fig:synallex}.

\begin{table}[htp!]
	\begin{center}
		\centering
		\caption{Comparison of different image synthesis methods in terms of DICE (segmentation performance of CBAS when trained on a mixture of real and synthetic images, synthesized using methods which are compared)and SSIM measures.}
		\resizebox{0.3\textwidth}{!}{
			\label{Comp_syn}
			\begin{tabular}{c|c|c}
				\hline
				Method       & DICE    & SSIM \\ \hline 
				pix2pix\cite{isola2017image}        & 0.8012   & 0.1995  \\ \hline
				pix2pixHD\cite{wang2018high}        & 0.8623   & 0.2643  \\ \hline
				MSSA(ours) & \textbf{0.8901}  & \textbf{0.2759}  
				\\  \hline
			\end{tabular}
		}
	\end{center}
\end{table}

\begin{figure}[htp!]
	\begin{center}
		\centering
		\includegraphics[width=0.52\textwidth]{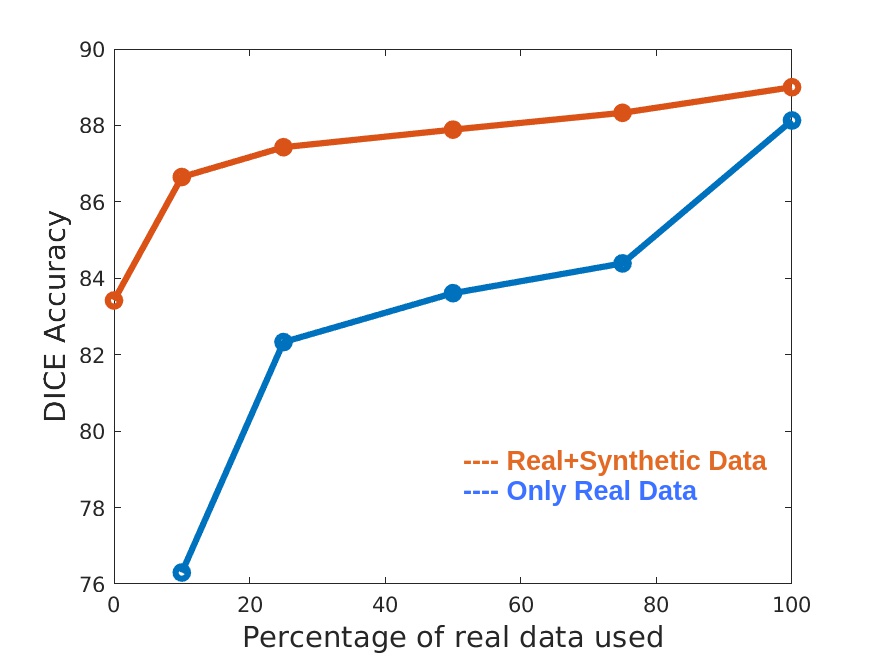}\\
		\caption{Performance of the segmentation network while trained on different proportions of the real data.}
		\label{Fig:percent}
	\end{center}
\end{figure}

We conduct further experiments to ascertain the importance of the synthetic data that is generated. Table ~\ref{Comp_per} contains DICE accuracies of the CBAS network when trained with different proportions of the real data. The total number of images on which the images are trained are always 1300 in every case. The percentage of real data out of the 1300 is different for every case except for the 100\% case. For example when the network is trained with 50\% real data, it is trained with 650 images. When it is trained with 50\% real and synthetic data, it is trained with 650 real images and 650 synthetic images. Only in the 100\% case, the number of images used for the real case is 1300 and the number of images used for real with synthetic data is 2600 images. It can be seen from the table that the addition of synthetic data is highly useful in cases where the real data availability is very low. Also, even when the network is trained only on the synthetic data, it gives a dice accuracy of 83.42\%. We also illustrate the performance gap produced by adding synthetic data in Fig \ref{Fig:percent} where we can see how synthetic data helps when the real data is very less. It should also be noted that in all these experiments (except in 100\% case), the upper bound of the number of training data was fixed as 1300 to give a reasonable comparison. 

%However, many combinations of meaning manipulations can be done on the segmentation mask resulting in numerous synthesized images.  

\begin{table*}[htp!]
	\begin{center}
		\centering
		\caption{Comparison of the segmentation performance of CBAS across different proportions of the real data.}
		\resizebox{0.6\textwidth}{!}{
			\label{Comp_per}
			\begin{tabular}{c|c|c}
				\hline
				\hline
				\% Real       & DICE- CBAS (only real)  & DICE- CBAS (real+synthetic)  \\ \hline
				100 \%        & 0.8813 & 0.8901   \\ \hline
				75 \%        & 0.8439  & 0.8833\\ \hline
				50 \%        & 0.8361  & 0.8789\\ \hline
				25 \%        & 0.8233  & 0.8743\\ \hline
				10 \%        & 0.7630  & 0.8665\\ \hline
				
				0 \% (only synthetic) & - & 0.8342 \\ \hline \hline
			\end{tabular}
		}
	\end{center}
\end{table*}

\subsection{Ablation Study}
We study the performance of each block's contribution to CBAS by conducting various experiments on the test images. We start with the UNet base network (BN), and then add SB blocks to estimate the segmentation maps at different scales. Finally, we add CB block to construct CBAS and train it with $\mathcal{L}_{final}$. Table.~\ref{ABL} shows the contribution of each block on the CBAS network. Note that BN and BN w/ SB are trained using the cross-entropy (CE) loss. The base network, BN itself produces poor results. However, when SB blocks are added to BN, the performance improves significantly. The combination of BN, SB and CB to construct CBAS and trained with $\mathcal{L}_{final}$ produces the best results. Table.~\ref{ABL} clearly shows the importance of formulating ML inference and training CBAS with $\mathcal{L}_{final}$.  This can be clearly seen by comparing the performance of CBAS when trained with and without $\mathcal{L}_{final}$. We computed the $p$ values using the DICE scores for the results obtained after adding different components to base network to obtain CBAS, against the DICE scores for the final results obtained using CBAS shown in the Table.~\ref{ABL}. 

Fig.~\ref{Fig:exp3} shows the qualitative performance of BN, BN w/SB, and CBAS. We can clearly see the progressive improvements visually when each block is added to BN. For example in the first column of Fig.~\ref{Fig:exp3}, the output brain ventricular segmentation regions are random at the edges for large size ventricles, and contains over segmentation of normal regions for small size ventricles. Once we add the SB blocks to BN, the outputs get much better compared to BN, but we can still observe some under segmentations in large ventricles and over segmentation in small size ventricles as shown in the third column of Fig~\ref{Fig:exp3}. Finally, when we add the CB blocks to construct CBAS and train it with $\mathcal{L}_{final}$, we observe the best results as shown in the fourth column of Fig.~\ref{Fig:exp3}.  Final outputs have clear edges for larger ventricles and accurate segmentation for smaller size ventricles.

Fig.~\ref{Fig:exp4} shows brain ventricle segmentation, and the corresponding confidence maps at different scales. We clearly observe $c_{\times 2}$, $c_{\times 1}$ (fourth and sixth columns in Fig.~\ref{Fig:exp4} respectively) highlight the erroneous regions $\hat{s}_{\times 2}$, $\hat{s}_{\times 1}$ (third and fifth columns in Fig.~\ref{Fig:exp4} respectively) which guide the CBAS to learn the accurate segmentation in those regions. For example, as shown in  Fig.~\ref{Fig:exp4}, the edges of the brain ventricle segmentation are highlighted in the confidence maps by producing low confidence scores using the CB blocks. This makes CBAS more attentive in those regions while calculating the segmentation maps.

\section{Conclusion}
\label{conc}
We proposed a novel method, called CBAS, to address the US brain anatomy segmentation task. In our approach, we introduced a technique to estimate segmentation and the corresponding confidence maps. Additionally, we trained our CBAS network with proposed novel loss function $\mathcal{L}_{final}$. Extensive experiments showed that CBAS outperformed the state-of-the-art methods with fewer number of parameters. The reported computational time makes CBAS the best match for real-time applications. On top of that, we proposed a image synthesis method to add synthetic data to our training data, which further boosts the performance of CBAS. We also show from various experiments that our proposed synthesis method is better than recent methods.

Although, our proposed method outperforms state-of-the-art methods, several limitations in our study still exists. First our method is geared towards segmenting 2D US data. 2D scans are inherently limited to cross-sectional analysis and do not take advantage of surface continuity between adjacent images (i.e., along the axis perpendicular to the scan plane direction). Currently, we are in the process of collecting 3D US scans. Therefore, in the future, we will extend our method for processing volumetric US data. Second limitation is related to the the fact that manual segmentation, performed by single expert ultrasonographer with more than 20 years of experience, was treated as gold standard in our study. Due to the typical US imaging artifacts manual segmentation of US data is an error prone process. Shape of the anatomical region to be segmented and expertise of the ultrasonographer will bias the obtained segmentation results. Future work will also involve the investigation of inter- and intra-user variability of the segmentation and its effect on the proposed method. During monitoring of the preterm neonates in situations where the diagnosis can not be assessed with US additional imaging using MRI is performed. Anatomical structures segmented from MRI data could be treated as a gold standard segmentation to minimize the variability of manual segmentation from US data. Unfortunately, none of the enrolled subjects had an MRI scan available. Therefore this analysis could not be performed during this work. We also did not calculate any quantitative US measurements such as ventricular index (VI), anterior horn width (AHW), and thalamo-occipital distance (TOD). These measurements are usually calculated manually from B-mode US data \cite{davies2000reference}. In or future work we will extend out network for simultaneous segmentation and anatomical landmark extraction in order to automate the quantitative measurement process. Finally, during this work we have only focused on lateral ventricles and septum pellecudi. Segmentation of third and fourth ventricles were beyond the scope of this study. However, quantitative measures obtained from these ventricles should be considered as a valuable additional information to evaluate the pathophysiology of ventriculomegaly \cite{brouwer2012new}.

\appendices
\section{Details of different blocks in CBAS}
Table \ref{block_tables} shows the details regarding ResBlock, Segmentation Block and Confidence Block in our network. Note that in Table~\ref{block_tables} $C, H$ and $W$ denote the number of channels, height and width of the intermediate feature maps respectively.
\begin{table}[htp!]
	\begin{center}
		\centering
		\caption{Configuration of blocks in the CBAS network.}
		\label{block_tables}
		\resizebox{0.48\textwidth}{!}{
			\begin{tabular}{c|c|c|c|c|c|c}
				\hline \hline
				Block name                          & Layer     & Kernel size & Filters & dilation & Input size & Output size \\ \hline
				\multirow{3}{*}{ResBlock}           & Conv1     & 1 x 1       & 2C      & 1          & C $\times$ H $\times$ W  & 2C $\times$ H $\times$ W  \\ \cline{2-7} 
				& Conv2     & 3 $\times$ 3       & 2C      & 1          & 2C $\times$ H $\times$ W & 2C $\times$ H $\times$ W  \\ \cline{2-7} 
				& Conv3     & 3 $\times$ 3       & C       & 2          & 2C $\times$ H $\times$ W & C $\times$ H $\times$ W   \\ \hline
				\multirow{4}{*}{Segmentation Block} & Conv1     & 1 $\times$ 1       & 32      & 1          & 64 $\times$ H $\times$ W & 32 $\times$ H $\times$ W  \\ \cline{2-7} 
				& Conv2     & 3 $\times$ 3       & 32      & 1          & 32 $\times$ H $\times$ W & 32 $\times$ H $\times$ W  \\ \cline{2-7} 
				& Conv3     & 3 $\times$ 3       &16     & 1          & 32 $\times$ H $\times$ W & 16 $\times$ H $\times$ W  \\ \cline{2-7} 
				& Conv4     & 3 $\times$ 3       & 1       & 1          & 16 $\times$ H $\times$ W & 1 $\times$ H $\times$ W   \\ \hline
				\multirow{5}{*}{Confidence Block}   & Conv1     & 1 $\times$ 1       & 16      & 1          & 33 $\times$ H $\times$ W & 16 $\times$ H $\times$ W  \\ \cline{2-7} 
				& Conv2     & 3 $\times$ 3       & 16      & 1          & 16 $\times$ H $\times$ W & 16 $\times$ H $\times$ W  \\ \cline{2-7} 
				& Conv3     & 3 $\times$ 3       & 16      & 1          & 16 $\times$ H $\times$ W & 16 $\times$ H $\times$ W  \\ \cline{2-7} 
				& Conv4     & 3 $\times$ 3       & 1       & 1          & 16 $\times$ H $\times$ W & 1 $\times$ H $\times$ W   \\ \cline{2-7} 
				& Sigmoid & --          & --      & --         & 1 $\times$ H $\times$ W  & 1 $\times$ H $\times$ W   \\ \hline \hline
			\end{tabular}
		}
	\end{center}
\end{table}

\section{Details of MSSA Network}

\subsection{Generator}
Table~\ref{block_tables_g} shows the details of each block in the generator network's architecture. Note that, $k$ is the number of filters in the convolutional layers in blocks, where ever specified. $C$ is the number of channels of input fed into the convolutional layer in the blocks, where ever specified.

\begin{table}[htp!]
	\begin{center}
		\centering
		\caption{Configuration of the synthesis generator network.}
		\label{block_tables_g}
		\resizebox{0.48\textwidth}{!}{
			\begin{tabular}{c|c|c|c|c|c|c}
				\hline \hline
				Block name                          & Layer     & Kernel size & Filters & Stride & Input size & Output size \\ \hline
				\multirow{3}{*}{ConvBlock 1}           & Conv1     & 7 $\times$ 7       & k     & 1          & 1 $\times$ H $\times$ W  & k $\times$ H $\times$ W  \\ \cline{2-7} 
				& InstanceNorm     & --       & --      & --          & k $\times$ H $\times$ W & k $\times$ H $\times$ W  \\ \cline{2-7} 
				& ReLU     & --       & --       & --          & k $\times$ H $\times$ W & k $\times$ H $\times$ W   \\ \hline
				\multirow{4}{*}{ConvBlock 2} & Conv1     & 3 $\times$ 3       & k      & 2          & C $\times$ H $\times$ W & k $\times$ H/2 $\times$ W/2  \\ \cline{2-7} 
				& InstanceNorm     & --       & --      & --          & k $\times$ H/2 $\times$ W/2 & k $\times$ H/2 $\times$ W/2  \\ \cline{2-7} 
				& ReLU     & --       & --     & --          & k $\times$ H/2 $\times$ W/2 & k $\times$ H/2 $\times$ W/2   \\ \hline
				\multirow{3}{*}{ConvBlock 3}           & Conv1     & 3 x 3       & k      & 0.5          & C $\times$ H $\times$ W  & k $\times$ 2H $\times$ 2W  \\ \cline{2-7} 
				& InstanceNorm     & --       & --      & --          & k $\times$ 2H $\times$ 2W & k $\times$ 2H $\times$ 2W  \\ \cline{2-7} 
				& ReLU     & --       & --       & --          & k $\times$ 2H $\times$ 2W & k $\times$ 2H $\times$ 2W   \\ \hline
				\multirow{3}{*}{Self Attention Block}           & Query-Conv1     & 1 x 1       & 128      & 1          & C $\times$ H $\times$ W  & 128 $\times$ H $\times$ W  \\ \cline{2-7} 
				& Key-Conv2     & 1 $\times$ 1       & 128      & 1          & 128 $\times$ H $\times$ W & 128 $\times$ H $\times$ W  \\ \cline{2-7} 
				& Value-Conv3     & 1 $\times$ 1       & 1024       &           & 128 $\times$ H $\times$ W & 1024 $\times$ H $\times$ W   \\ \hline
				
				\multirow{2}{*}{ResBlock}   & Conv1     & 3 $\times$ 3       & k      & 1          & C $\times$ H $\times$ W & k $\times$ H $\times$ W  \\ \cline{2-7} 
				& Conv2     & 3 $\times$ 3       & k      & 1          & k $\times$ H $\times$ W & k $\times$ H $\times$ W   \\ \hline \hline
				
			\end{tabular}
		}
	\end{center}
\end{table}

\subsection{Discriminator}
Table~\ref{block_tables_d} shows the details of each block in the discriminator's network architecture. Note that, $k$ is the number of filters in the convolutional layers in the block.

\begin{table}[htp!]
	\begin{center}
		\centering
		\caption{Configuration of blocks in the discriminator network.}
		\label{block_tables_d}
		\resizebox{0.48\textwidth}{!}{
			\begin{tabular}{c|c|c|c|c|c|c}
				\hline \hline
				Block name                          & Layer     & Kernel size & Filters & Stride & Input size & Output size \\ \hline
				\multirow{3}{*}{ConvBlock}           & Conv1     & 4 x 4       & k      & 1          & 1 $\times$ H $\times$ W  & k $\times$ H $\times$ W  \\ \cline{2-7} 
				& InstanceNorm     & --       & --      & --          & k $\times$ H $\times$ W & k $\times$ H $\times$ W  \\ \cline{2-7} 
				& LeakyReLU     & --       & --       & --          & k $\times$ H $\times$ W & k $\times$ H $\times$ W   \\ \hline \hline
				 
			\end{tabular}
		}
	\end{center}
\end{table}

\bibliographystyle{IEEEtran}
\bibliography{myref}

% Generated by IEEEtran.bst, version: 1.14 (2015/08/26)
\begin{thebibliography}{10}
\providecommand{\url}[1]{#1}
\csname url@samestyle\endcsname
\providecommand{\newblock}{\relax}
\providecommand{\bibinfo}[2]{#2}
\providecommand{\BIBentrySTDinterwordspacing}{\spaceskip=0pt\relax}
\providecommand{\BIBentryALTinterwordstretchfactor}{4}
\providecommand{\BIBentryALTinterwordspacing}{\spaceskip=\fontdimen2\font plus
\BIBentryALTinterwordstretchfactor\fontdimen3\font minus
  \fontdimen4\font\relax}
\providecommand{\BIBforeignlanguage}[2]{{%
\expandafter\ifx\csname l@#1\endcsname\relax
\typeout{** WARNING: IEEEtran.bst: No hyphenation pattern has been}%
\typeout{** loaded for the language `#1'. Using the pattern for}%
\typeout{** the default language instead.}%
\else
\language=\csname l@#1\endcsname
\fi
#2}}
\providecommand{\BIBdecl}{\relax}
\BIBdecl

\bibitem{blencowe2013born}
H.~Blencowe, S.~Cousens, D.~Chou, M.~Oestergaard, L.~Say, A.-B. Moller,
  M.~Kinney, and J.~Lawn, ``Born too soon: the global epidemiology of 15
  million preterm births,'' \emph{Reproductive health}, vol.~10, no.~1, p.~S2,
  2013.

\bibitem{robinson2012neonatal}
S.~Robinson, ``Neonatal posthemorrhagic hydrocephalus from prematurity:
  pathophysiology and current treatment concepts: a review,'' \emph{Journal of
  Neurosurgery: Pediatrics}, vol.~9, no.~3, pp. 242--258, 2012.

\bibitem{sherer2004prenatal}
D.~M. Sherer, M.~Sokolovski, M.~Dalloul, P.~Santoso, J.~Curcio, and
  O.~Abulafia, ``Prenatal diagnosis of dilated cavum septum pellucidum et
  vergae,'' \emph{American journal of perinatology}, vol.~21, no.~05, pp.
  247--251, 2004.

\bibitem{sarwar1989septum}
M.~Sarwar, ``The septum pellucidum: normal and abnormal.'' \emph{American
  Journal of Neuroradiology}, vol.~10, no.~5, pp. 989--1005, 1989.

\bibitem{isola2017image}
P.~Isola, J.-Y. Zhu, T.~Zhou, and A.~A. Efros, ``Image-to-image translation
  with conditional adversarial networks,'' in \emph{Proceedings of the IEEE
  conference on computer vision and pattern recognition}, 2017, pp. 1125--1134.

\bibitem{unet}
O.~Ronneberger, P.~Fischer, and T.~Brox, ``U-net: Convolutional networks for
  biomedical image segmentation,'' in \emph{International Conference on Medical
  Image Computing and Computer-Assisted Intervention}.\hskip 1em plus 0.5em
  minus 0.4em\relax Springer, 2015, pp. 234--241.

\bibitem{wang2018automatic}
P.~Wang, N.~G. Cuccolo, R.~Tyagi, I.~Hacihaliloglu, and V.~M. Patel,
  ``Automatic real-time cnn-based neonatal brain ventricles segmentation,'' in
  \emph{2018 IEEE 15th International Symposium on Biomedical Imaging (ISBI
  2018)}.\hskip 1em plus 0.5em minus 0.4em\relax IEEE, 2018, pp. 716--719.

\bibitem{wang2018high}
T.-C. Wang, M.-Y. Liu, J.-Y. Zhu, A.~Tao, J.~Kautz, and B.~Catanzaro,
  ``High-resolution image synthesis and semantic manipulation with conditional
  gans,'' in \emph{Proceedings of the IEEE conference on computer vision and
  pattern recognition}, 2018, pp. 8798--8807.

\bibitem{qiu2017automatic}
W.~Qiu, Y.~Chen, J.~Kishimoto, S.~de~Ribaupierre, B.~Chiu, A.~Fenster, and
  J.~Yuan, ``Automatic segmentation approach to extracting neonatal cerebral
  ventricles from 3d ultrasound images,'' \emph{Medical image analysis},
  vol.~35, pp. 181--191, 2017.

\bibitem{boucher2018dilatation}
M.-A. Boucher, S.~Lipp{\'e}, A.~Damphousse, R.~El-Jalbout, and S.~Kadoury,
  ``Dilatation of lateral ventricles with brain volumes in infants with 3d
  transfontanelle us,'' in \emph{International Conference on Medical Image
  Computing and Computer-Assisted Intervention}.\hskip 1em plus 0.5em minus
  0.4em\relax Springer, 2018, pp. 557--565.

\bibitem{sciolla2016segmentation}
B.~Sciolla, M.~Martin, P.~Delachartre, and P.~Quetin, ``Segmentation of the
  lateral ventricles in 3d ultrasound images of the brain in neonates,'' in
  \emph{2016 IEEE International Ultrasonics Symposium (IUS)}.\hskip 1em plus
  0.5em minus 0.4em\relax IEEE, 2016, pp. 1--4.

\bibitem{martin2018automatic}
M.~Martin, B.~Sciolla, M.~Sdika, X.~Wang, P.~Quetin, and P.~Delachartre,
  ``Automatic segmentation of the cerebral ventricle in neonates using deep
  learning with 3d reconstructed freehand ultrasound imaging,'' in \emph{2018
  IEEE International Ultrasonics Symposium (IUS)}.\hskip 1em plus 0.5em minus
  0.4em\relax IEEE, 2018, pp. 1--4.

\bibitem{hamaguchi2018effective}
R.~Hamaguchi, A.~Fujita, K.~Nemoto, T.~Imaizumi, and S.~Hikosaka, ``Effective
  use of dilated convolutions for segmenting small object instances in remote
  sensing imagery,'' in \emph{2018 IEEE Winter Conference on Applications of
  Computer Vision (WACV)}.\hskip 1em plus 0.5em minus 0.4em\relax IEEE, 2018,
  pp. 1442--1450.

\bibitem{KendallGal2017UncertaintiesB}
A.~Kendall and Y.~Gal, ``{What Uncertainties Do We Need in Bayesian Deep
  Learning for Computer Vision?}'' in \emph{Advances in Neural Information
  Processing Systems 30 (NIPS)}, 2017.

\bibitem{kendall2017multi}
A.~Kendall, Y.~Gal, and R.~Cipolla, ``Multi-task learning using uncertainty to
  weigh losses for scene geometry and semantics,'' in \emph{Proceedings of the
  IEEE Conference on Computer Vision and Pattern Recognition ({CVPR})}, 2018.

\bibitem{mehta2019propagating}
R.~Mehta, T.~Christinck, T.~Nair, P.~Lemaitre, D.~Arnold, and T.~Arbel,
  ``Propagating uncertainty across cascaded medical imaging tasks for improved
  deep learning inference,'' in \emph{Uncertainty for Safe Utilization of
  Machine Learning in Medical Imaging and Clinical Image-Based
  Procedures}.\hskip 1em plus 0.5em minus 0.4em\relax Springer, 2019, pp.
  23--32.

\bibitem{nair2020exploring}
T.~Nair, D.~Precup, D.~L. Arnold, and T.~Arbel, ``Exploring uncertainty
  measures in deep networks for multiple sclerosis lesion detection and
  segmentation,'' \emph{Medical image analysis}, vol.~59, p. 101557, 2020.

\bibitem{jungo2019assessing}
A.~Jungo and M.~Reyes, ``Assessing reliability and challenges of uncertainty
  estimations for medical image segmentation,'' in \emph{International
  Conference on Medical Image Computing and Computer-Assisted
  Intervention}.\hskip 1em plus 0.5em minus 0.4em\relax Springer, 2019, pp.
  48--56.

\bibitem{goodfellow2014generative}
I.~Goodfellow, J.~Pouget-Abadie, M.~Mirza, B.~Xu, D.~Warde-Farley, S.~Ozair,
  A.~Courville, and Y.~Bengio, ``Generative adversarial nets,'' in
  \emph{Advances in neural information processing systems}, 2014, pp.
  2672--2680.

\bibitem{mirza2014conditional}
M.~Mirza and S.~Osindero, ``Conditional generative adversarial nets,''
  \emph{arXiv preprint arXiv:1411.1784}, 2014.

\bibitem{nie2018medical}
D.~Nie, R.~Trullo, J.~Lian, L.~Wang, C.~Petitjean, S.~Ruan, Q.~Wang, and
  D.~Shen, ``Medical image synthesis with deep convolutional adversarial
  networks,'' \emph{IEEE Transactions on Biomedical Engineering}, vol.~65,
  no.~12, pp. 2720--2730, 2018.

\bibitem{nie2017medical}
D.~Nie, R.~Trullo, J.~Lian, C.~Petitjean, S.~Ruan, Q.~Wang, and D.~Shen,
  ``Medical image synthesis with context-aware generative adversarial
  networks,'' in \emph{International Conference on Medical Image Computing and
  Computer-Assisted Intervention}.\hskip 1em plus 0.5em minus 0.4em\relax
  Springer, 2017, pp. 417--425.

\bibitem{wolterink2017deep}
J.~M. Wolterink, A.~M. Dinkla, M.~H. Savenije, P.~R. Seevinck, C.~A. van~den
  Berg, and I.~I{\v{s}}gum, ``Deep mr to ct synthesis using unpaired data,'' in
  \emph{International Workshop on Simulation and Synthesis in Medical
  Imaging}.\hskip 1em plus 0.5em minus 0.4em\relax Springer, 2017, pp. 14--23.

\bibitem{bi2017synthesis}
L.~Bi, J.~Kim, A.~Kumar, D.~Feng, and M.~Fulham, ``Synthesis of positron
  emission tomography (pet) images via multi-channel generative adversarial
  networks (gans),'' in \emph{Molecular Imaging, Reconstruction and Analysis of
  Moving Body Organs, and Stroke Imaging and Treatment}.\hskip 1em plus 0.5em
  minus 0.4em\relax Springer, 2017, pp. 43--51.

\bibitem{han2018gan}
C.~Han, H.~Hayashi, L.~Rundo, R.~Araki, W.~Shimoda, S.~Muramatsu, Y.~Furukawa,
  G.~Mauri, and H.~Nakayama, ``Gan-based synthetic brain mr image generation,''
  in \emph{2018 IEEE 15th International Symposium on Biomedical Imaging (ISBI
  2018)}.\hskip 1em plus 0.5em minus 0.4em\relax IEEE, 2018, pp. 734--738.

\bibitem{yang2018mri}
Q.~Yang, N.~Li, Z.~Zhao, X.~Fan, E.-C. Chang, Y.~Xu \emph{et~al.}, ``Mri
  image-to-image translation for cross-modality image registration and
  segmentation,'' \emph{arXiv preprint arXiv:1801.06940}, 2018.

\bibitem{armanious2019medgan}
K.~Armanious, C.~Jiang, M.~Fischer, T.~K{\"u}stner, T.~Hepp, K.~Nikolaou,
  S.~Gatidis, and B.~Yang, ``Medgan: Medical image translation using gans,''
  \emph{Computerized Medical Imaging and Graphics}, p. 101684, 2019.

\bibitem{zhao2018synthesizing}
H.~Zhao, H.~Li, S.~Maurer-Stroh, and L.~Cheng, ``Synthesizing retinal and
  neuronal images with generative adversarial nets,'' \emph{Medical image
  analysis}, vol.~49, pp. 14--26, 2018.

\bibitem{bailo2019red}
O.~Bailo, D.~Ham, and Y.~Min~Shin, ``Red blood cell image generation for data
  augmentation using conditional generative adversarial networks,'' in
  \emph{Proceedings of the IEEE Conference on Computer Vision and Pattern
  Recognition Workshops}, 2019, pp. 0--0.

\bibitem{jaiswal2018capsulegan}
A.~Jaiswal, W.~AbdAlmageed, Y.~Wu, and P.~Natarajan, ``Capsulegan: Generative
  adversarial capsule network,'' in \emph{Proceedings of the European
  Conference on Computer Vision (ECCV)}, 2018, pp. 0--0.

\bibitem{fujioka2019breast}
T.~Fujioka, M.~Mori, K.~Kubota, Y.~Kikuchi, L.~Katsuta, M.~Adachi, G.~Oda,
  T.~Nakagawa, Y.~Kitazume, and U.~Tateishi, ``Breast ultrasound image
  synthesis using deep convolutional generative adversarial networks,''
  \emph{Diagnostics}, vol.~9, no.~4, p. 176, 2019.

\bibitem{radford2015unsupervised}
A.~Radford, L.~Metz, and S.~Chintala, ``Unsupervised representation learning
  with deep convolutional generative adversarial networks,'' \emph{arXiv
  preprint arXiv:1511.06434}, 2015.

\bibitem{hu2017freehand}
Y.~Hu, E.~Gibson, L.-L. Lee, W.~Xie, D.~C. Barratt, T.~Vercauteren, and J.~A.
  Noble, ``Freehand ultrasound image simulation with spatially-conditioned
  generative adversarial networks,'' in \emph{Molecular imaging, reconstruction
  and analysis of moving body organs, and stroke imaging and treatment}.\hskip
  1em plus 0.5em minus 0.4em\relax Springer, 2017, pp. 105--115.

\bibitem{tom2018simulating}
F.~Tom and D.~Sheet, ``Simulating patho-realistic ultrasound images using deep
  generative networks with adversarial learning,'' in \emph{2018 IEEE 15th
  International Symposium on Biomedical Imaging (ISBI 2018)}.\hskip 1em plus
  0.5em minus 0.4em\relax IEEE, 2018, pp. 1174--1177.

\bibitem{Instance2016}
D.~Ulyanov, A.~Vedaldi, and V.~Lempitsky, ``Instance normalization: The missing
  ingredient for fast stylization.''

\bibitem{huang2017densely}
G.~Huang, Z.~Liu, L.~van~der Maaten, and K.~Q. Weinberger, ``Densely connected
  convolutional networks,'' in \emph{Proceedings of the IEEE Conference on
  Computer Vision and Pattern Recognition}, 2017.

\bibitem{zhang2018self}
H.~Zhang, I.~Goodfellow, D.~Metaxas, and A.~Odena, ``Self-attention generative
  adversarial networks,'' \emph{arXiv preprint arXiv:1805.08318}, 2018.

\bibitem{vaswani2017attention}
A.~Vaswani, N.~Shazeer, N.~Parmar, J.~Uszkoreit, L.~Jones, A.~N. Gomez,
  {\L}.~Kaiser, and I.~Polosukhin, ``Attention is all you need,'' in
  \emph{Advances in neural information processing systems}, 2017, pp.
  5998--6008.

\bibitem{he2016deep}
K.~He, X.~Zhang, S.~Ren, and J.~Sun, ``Deep residual learning for image
  recognition,'' in \emph{Proceedings of the IEEE conference on computer vision
  and pattern recognition}, 2016, pp. 770--778.

\bibitem{long2015fully}
J.~Long, E.~Shelhamer, and T.~Darrell, ``Fully convolutional networks for
  semantic segmentation,'' in \emph{Proceedings of the IEEE conference on
  computer vision and pattern recognition}, 2015, pp. 3431--3440.

\bibitem{johnson2016perceptual}
J.~Johnson, A.~Alahi, and L.~Fei-Fei, ``Perceptual losses for real-time style
  transfer and super-resolution,'' in \emph{European conference on computer
  vision}.\hskip 1em plus 0.5em minus 0.4em\relax Springer, 2016, pp. 694--711.

\bibitem{kingma2014adam}
D.~P. Kingma and J.~Ba, ``Adam: A method for stochastic optimization,''
  \emph{arXiv preprint arXiv:1412.6980}, 2014.

\bibitem{davies2000reference}
M.~Davies, M.~Swaminathan, S.~Chuang, and F.~Betheras, ``Reference ranges for
  the linear dimensions of the intracranial ventricles in preterm neonates,''
  \emph{Archives of Disease in Childhood-Fetal and Neonatal Edition}, vol.~82,
  no.~3, pp. F218--F223, 2000.

\bibitem{brouwer2012new}
M.~J. Brouwer, L.~S. de~Vries, F.~Groenendaal, C.~Koopman, L.~R. Pistorius,
  E.~J. Mulder, and M.~J. Benders, ``New reference values for the neonatal
  cerebral ventricles,'' \emph{Radiology}, vol. 262, no.~1, pp. 224--233, 2012.

\end{thebibliography}

\end{document}